\documentclass[fleqn,usenatbib]{mnras}

\usepackage{newtxtext,newtxmath}

\usepackage[T1]{fontenc}

\DeclareRobustCommand{\VAN}[3]{#2}
\let\VANthebibliography\thebibliography
\def\thebibliography{\DeclareRobustCommand{\VAN}[3]{##3}\VANthebibliography}

\usepackage{graphicx}	
\usepackage{amsmath}	

\newcommand{\Nsample}{75}
\newcommand{\re}{$R_{\rm e}$}
\newcommand{\br}{$\beta_{\rm Re}$}
\makeatletter
\newcommand\sendemail[3]{
\edef\@tempa{mailto:#1?subject=#2 }%
\edef\@tempb{\expandafter\html@spaces\@tempa\@empty}%
\href{\@tempb}{#3}}

\catcode\%=11
\def\html@spaces#1 #2{#1
\catcode\%=14
\makeatother

\title[]{The MAGPI Survey: Orbital distributions, intrinsic shapes, and mass profiles for MAGPI-like {\sc Eagle} galaxies using Schwarzschild dynamical models}

\author[G. Santucci et al.]{\parbox{\textwidth}{
Giulia Santucci,$^{1,2}$\thanks{E-mail: giulia.santucci@uwa.edu.au}
Claudia Del P. Lagos,$^{1,2}$
Katherine E. Harborne,$^{1,2}$
Caro Derkenne,$^{3,2}$
Adriano Poci,$^{4}$
Sabine Thater,$^{5}$
Richard M. McDermid,$^{3,2}$
J. Trevor Mendel, $^{6,2}$
Emily Wisnioski, $^{6.2}$
Scott M. Croom,$^{7,2}$
Anna Ferr\'e-Mateu$^{8,9}$
Eric G. M. Muller,$^{6,2}$
Jesse van de Sande,$^{10,2}$
Gauri Sharma,$^{11,12,13,14}$
Sarah M. Sweet, $^{15,2}$
Takafumi Tsukui,$^{6,2}$
Lucas M. Valenzuela, $^{16}$
Glenn van de Ven,$^{5}$
Tayyaba Zafar$^{3}$
}
\vspace{0.4cm}
\\
\parbox{\textwidth}{
$^{1}$ International Centre for Radio Astronomy Research (ICRAR), M468, University of Western Australia, 35 Stirling Hwy, Crawley, WA 6009, Australia\\
$^{2}$ARC Centre of Excellence for All Sky Astrophysics in 3 Dimensions (ASTRO 3D), Australia\\
$^{3}$Research Centre for Astronomy, Astrophysics, and Astrophotonics, School of Mathematical and Physical Sciences, Macquarie University,NSW 2109, Australia\\
$^{4}$Astrophysics Sub-department, Department of Physics, University of Oxford, Keble Road, Oxford OX1 3RH, United Kingdom\\
$^{5}$Department of Astrophysics, University of Vienna, T\"{u}rkenschanzstrasse 17, 1180 Wien, Austria\\
$^{6}$Research School of Astronomy and Astrophysics, Australian National University, Canberra, ACT 2611, Australia\\
$^{7}$Sydney Institute for Astronomy, School of Physics, University of Sydney, NSW 2006, Australia\\
$^{8}$Instituto de Astrof\'isica de Canarias, c/ V\'ia L\'actea s/n, E-38205 - La Laguna, Tenerife, Spain\\
$^{9}$Departamento de Astrof\'isica, Universidad de La Laguna, E-38205 - La Laguna, Tenerife, Spain\\
$^{10}$ School of Physics, University of New South Wales, NSW 2052, Australia\\
$^{11}$ Observatoire Astronomique de Strasbourg, Université de Strasbourg, CNRS UMR 7550, F-67000 Strasbourg, France\\
$^{12}$ University of Strasbourg Institute for Advanced Study, 5 allée du Général Rouvillois, F-67083 Strasbourg, France\\
$^{13}$Department of Physics and Astronomy, University of the Western Cape, Cape Town 7535, South Africa\\
$^{14}$ SISSA International School for Advanced Studies, Via Bonomea 265, I-34136 Trieste, Italy\\
$^{15}$School of Mathematics and Physics, University of Queensland, Brisbane, QLD 4072, Australia\\
$^{16}$Universit\"{a}ts-Sternwarte, Fakult\"{a}t f\"{u}r Physik,Ludwig-Maximilians-Universit\"{a}t M\"{u}nchen, Scheinerstr. 1, 81679 M\"{u}nchen, Germany\\
}}

\date{Accepted XXX. Received YYY; in original form ZZZ}

\pubyear{2023}

\begin{document}
\label{firstpage}
\pagerange{\pageref{firstpage}--\pageref{lastpage}}
\maketitle

\begin{abstract}
Schwarzschild dynamical models are now regularly employed in large surveys of galaxies in the local and distant Universe to derive information on galaxies’ intrinsic properties such as their orbital structure and their (dark matter and stellar) mass distribution. Comparing the internal orbital structures and mass distributions of galaxies in the distant Universe with simulations is key to understanding what physical processes are responsible for shaping galaxy properties. However it is first crucial to understand whether observationally derived properties are directly comparable with intrinsic ones in simulations. To assess this, we build Schwarzschild dynamical models for MUSE-like IFS cubes (constructed to be like those obtained by the MAGPI survey) of \Nsample~ galaxies at $z \sim$ 0.3 from the {\sc Eagle} simulations. We compare the true particle-derived properties with the galaxies' model-derived properties. In general, we find that the models can recover the true galaxy properties qualitatively well, with the exception of the enclosed dark matter, where we find a median offset of 48\%, which is due to the assumed NFW profile not being able to reproduce the dark matter distribution in the inner region of the galaxies. We then compare our model-derived properties with Schwarzschild models-derived properties of observed MAGPI galaxies and find good agreement between MAGPI and {\sc Eagle}: the majority of our galaxies (57\%) have non-oblate shapes within 1 effective radius. More triaxial galaxies show higher fractions of hot orbits in their inner regions and tend to be more radially anisotropic.

\end{abstract}




\section{Introduction}
Over the last two decades, the rapidly growing field of Integral Field Spectroscopy (IFS) has significantly deepened our knowledge of stellar structures in galaxies in the local Universe, across a wide range of masses, morphologies, and environments. IFS instruments are the only ones that allow for the stellar and gas phase properties of galaxies to be spatially and simultaneously mapped. IFS surveys, such as ATLAS$^{\rm 3D}$ \citep{Cappellari2011}, the Sydney-AAO Multi-Object Integral-Field Spectrograph (SAMI) Galaxy Survey \citep{Croom2012, Bryant2015}, the Calar Alto Legacy Integral Field Area Survey (CALIFA; \citealt{Sanchez2012}), MASSIVE \citep{Ma2014} and the Mapping Nearby Galaxies at Apache Point Observatory (MaNGA) survey \citep{Bundy2015}, have contributed to significantly expand our understanding of galaxy kinematics and their connection to intrinsic galaxy properties and their environment \citep[see review by ][]{Cappellari2016}. For example, it has been possible to unveil various correlations between the proxy for the stellar spin parameter, $\lambda_{\rm Re}$, which provides a measurement of how rotationally supported a galaxy is, and different galaxy properties. $\lambda_{r}$ has previously been used to separate slow-rotating galaxies (i.e. galaxies whose kinematics are dominated by random motions) from fast-rotating galaxies \citep{Emsellem2007, Emsellem2011, Cappellari2016} and has been observed to be strongly correlated with stellar mass, so that the fraction of galaxies with low $\lambda_{\rm Re}$ (slow-rotating systems) increases with increasing stellar mass \citep{Emsellem2011,Falcon-Barroso2011, vandeSande2017a, Veale2017, Brough2017, Wang2020}, as well as correlated with environment \citep{vandeSande2021b}, and age \citep{vandeSande2018, Croom2024}. 

Thanks to these IFS surveys, kinematic maps as well as stellar age and metallicity maps for thousands of nearby galaxies have been made available. By applying orbit-based dynamical models to the IFS data \citep[e.g.][hereafter D24]{Cappellari2007, vandenBosch2008, Zhu2018hubble, Jin2020, Santucci2022, Santucci2023, Thater2023, Derkenne2024}, we can separate dynamically-derived properties of galaxies, such as intrinsic shape, orbital components, velocity anisotropy, and inner mass distribution. These dynamical models can therefore provide insight into the three-dimensional internal kinematic structures underlying the line-of-sight kinematics. Several different implementations of the Schwarzschild method, with varying degrees of symmetry, have been described \citep[e.g.][]{Cretton1999,Gebhardt2003, Valluri2004,vandenBosch2008, Vasiliev2015, Vasiliev2020, Neureiter2021}. The Schwarzschild method has been used to study the internal stellar structure of globular clusters \citep{vandeVen2006}, nuclear star clusters \citep{Feldmeier2017, Fahrion2019, Thater2023a}, and galaxies \citep[e.g.][]{Thomas2007,Cappellari2007, vandeVen2008, Feldmeier2017, Poci2019,Jin2020,Santucci2022,Pilawa2022, Thater2023,Santucci2023, Mehrgan2024, Jin2024agescalifa}, to measure the mass of supermassive black holes \citep{vanderMarel1998, Verolme2002,Gebhardt2003, Valluri2004, Krajnovic2009, Rusli2013, Seth2014, Thater2017, Thater2019, Liepold2020, Quenneville2021, Quenneville2021b, Thater2022bh, Neureiter2023, deNicola2024}, and has also been used to identify accreted galactic components \citep[e.g.][]{Zhu2020,Poci2021,Zhu2022halo}. 

However, to obtain a complete picture of galaxy evolution and understand how galaxies evolve into slow- and fast-rotating systems we also need observations at higher redshifts. The Middle Ages Galaxy Properties with Integral Field Spectroscopy (MAGPI) survey \citep{Foster2021} provides observations for a substantial sample of galaxies with both stellar and gas-phase observations at redshifts $z \sim 0.3$, an epoch previously only sparsely studied. The absence of substantial IFS observations in this epoch has limited our understanding of galaxy evolution during the Universe’s middle ages when morphology, angular momentum, and star formation activity evolve rapidly, with the environment potentially playing a key role \cite[e.g.][]{Peng2010, Choi2018}. The higher-redshift MAGPI sample, with its wide range in environment, can allow us to better constrain the internal properties of massive galaxies, and shed light on their later accretion history. In particular, thanks to the MAGPI theoretical library (Harborne et al., in prep), a like-to-like comparison with simulations is now possible. D24 built Schwarzschild dynamical models for 22 galaxies in the MAGPI survey. They found that the inner mass distribution of galaxies tends to be non-homogeneous, varying both between galaxies and within galaxies. They also found that the majority of the galaxies in their sample (over 85\%) had non-oblate shapes, and that the shape of a galaxy strongly correlates with the fraction of hot orbits within one effective radius.

MAGPI theoretical library (Harborne et al., in prep) provides the mock data from galaxy evolution simulations cubes with the same observational conditions and post-process of the MAGPI data. This enables us to assess the robustness of our modelling methods using simulated galaxies with ground truth. Such comparison will provide insights into how stellar orbital properties and mass distribution have evolved as simulations provide a mean to trace the evolutionary paths of galaxies over cosmic time.

In this paper, we build orbit-based triaxial dynamical model fits of \Nsample~MAGPI-like mock-observations of galaxies from the {\sc Eagle} simulations. We first compare the model-derived properties with the true properties, and then with the results from D24. In Sec. 2 we describe the sample of galaxies and the data available for this analysis; we describe how we construct the Schwarzschild dynamical models in Sec. 3; we present our results in Sec. 4 and we discuss them in Sec. 5. Our conclusions are given in Sec. 6. 

Throughout the paper, we adopt a $\Lambda CDM$ cosmology with $\Omega_m=0.3175$ and $H_0 = 67.1\ \mathrm{km\,s}^{-1} \mathrm{Mpc}^{-1}$, corresponding to the Planck 2013 cosmology \citep{Plank2014}.

\section{Data}
In this work we use the Schwarzschild modelling tool {\sc{DYNAMITE}} \citep{DYN2020, Thater2022dyna}. Our aim in this paper is first to test the robustness of its derived properties. Secondly, we aim to compare the properties derived for simulated galaxies with those obtained for MAGPI galaxies. For a fair comparison, we use the {\sc Eagle} - MAGPI theoretical library from the MAGPI Theory Data Release (Harborne et al. in prep). This library contains kinematic maps of MAGPI-like galaxies that can be analysed with observational methods. We introduce the {\sc Eagle} simulations in Sec. \ref{sec:eagledata}, the MAGPI theoretical library in Sec. \ref{sec:simspin}, the properties derived from the particle data in Sec. \ref{section:mass}, \ref{section:orbits}, and the sample selection in Sec. \ref{sec:selection}.

\subsection{The {\sc Eagle} simulation} \label{sec:eagledata}

The {\sc Eagle} simulations suite (\citealt{Schaye2015}, hereafter S15; \citealt{Crain2015}) employs a modified version of the N-body Tree-Particle-Mesh smoothed particle hydrodynamics (SPH) code GADGET-3 \citep{Springel2005Gadget} with subgrid recipes for radiative cooling, star formation, stellar evolution, chemical enrichment, stellar and black hole feedback. The suite consists of a large number of cosmological hydrodynamic simulations with different resolutions, volumes, and subgrid physics models, adopting the \cite{Plank2014} cosmology. S15 introduced a reference model, within which the parameters of the sub-grid models governing energy feedback from stars and accreting black holes were calibrated to ensure a good match to the $z = 0.1$ galaxy stellar mass function, the size-mass relation of present-day star-forming galaxies and the black hole-stellar mass relation (see \citealt{Crain2015} for details). In this work, we take the largest volume with the highest resolution (Ref-L100N1504) presented in S15. Table~\ref{TableSimus} summarises the parameters of this simulation.

\begin{table}
\begin{center}
  \caption{Specifications of the {\sc Eagle} Ref-L100N1504 simulation used in this paper. The rows list:
    (1) initial particle masses of gas and (2) dark matter, (3) the average stellar particle mass at birth and (4) at the redshift of interest $z \sim$ 0.27, (5) comoving Plummer-equivalent gravitational softening length, (6) maximum physical gravitational softening length. Units are indicated in each row. {\sc Eagle} adopts (5) as the softening length at $z\ge 2.8$, and (6) at $z<2.8$. This simulation has a side length of $L=100$~$\rm cMpc$. Here, pkpc and ckpc refer to proper and comoving kpc, respectively, and cMpc to comoving Mpc. }\label{TableSimus}
\begin{tabular}{l l l l}
\\[3pt]
\hline
& Property & Units & Value \\
\hline
(1)& gas particle mass & $[\rm M_{\odot}]$ & $1.81\times 10^6$\\
(2)& dark matter (DM) particle mass & $[\rm M_{\odot}]$ & $9.7\times 10^6$\\
(3)& average stellar particle mass at birth & $[\rm M_{\odot}]$ & $2.26\times 10^6$\\
(4)& average stellar particle mass at $z\sim 0.27$ & $[\rm M_{\odot}]$ & $1.3\times 10^6$\\
(5)& Softening length & $[\rm ckpc]$ & $2.66$\\
(6)& max. gravitational softening & $[\rm pkpc]$& $0.7$ \\
\end{tabular}
\end{center}
\end{table}

A major aspect of the {\sc Eagle} project is the use of state-of-the-art sub-grid models that capture unresolved physics (i.e. processes happening below the resolution limit). The sub-grid physics modules adopted by {\sc Eagle} are: (i) radiative cooling and photoheating \citep{Wiersma2009}, (ii) star formation \citep{Schaye2008}, (iii) stellar evolution and chemical enrichment \citep{Wiersma2009b}, (iv) stellar feedback \citep{DallaVecchia2012}, and (v) black hole growth and active galactic nucleus feedback 
\citep{Rosas-Guevara2015}. 

The {\sc Eagle} simulations were performed using an extensively modified version of GADGET-3 \citep{Springel2005Gadget}. Among those modifications are updates to the SPH technique, which are collectively referred to as ``Anarchy'' (see \citealt{Schaller2015} for an analysis of the impact of these changes on the properties of simulated galaxies compared to standard SPH). We use SUBFIND \citep{Springel2001,Dolag2009} to identify self-bound overdensities of particles within halos (i.e. substructures). These substructures are the galaxies in {\sc Eagle}.

Since we are interested in evaluating the robustness of the derived properties from Schwarzschild dynamical models for galaxies in the MAGPI survey, we use snapshot 25 at $z$ = 0.27, which is closest to the nominal 0.3 target redshift of the MAGPI survey. Due to the high computational time required for the models to be computed (around 400 CPU hours per galaxy, on average), we select a subsample of {\sc Eagle} galaxies in order to match the MAGPI Primary Targets (described in Sec. \ref{sec:simspin}). For each of these galaxies we follow the steps outlined below:
\begin{itemize}
   
    \item we calculate the true values for each of the galaxy properties we are interested in from the simulation stellar particle data (Sec. \ref{section:mass} and Sec. \ref{section:orbits}),
    \item for each mock observation, we emulate the process of fitting a dynamical model as for real data, by:
    \begin{enumerate}
        \item build Schwarzschild dynamical models using {\sc{DYNAMITE}},
        \item select the model that best matches our data, using a $\chi^2$ comparison,
        \item from the best-fit model, we derive the galaxy properties we are interested in (such as enclosed matter, intrinsic shape and orbital distribution),
    \end{enumerate}
    \item we compare the model-derived properties with the properties inferred from the particle data,
    \item we compare the model-derived properties with the properties derived for MAGPI galaxies by D24.

\end{itemize}

\subsection{The MAGPI theoretical library}\label{sec:simspin}
A suite of cosmological simulations is included in the MAGPI dataset. The MAGPI Survey uses existing cosmological hydrodynamical simulations and retrieves data from \textsc{Eagle} \citep{Schaye2015, Crain2015}, \textsc{Magneticum} \citep{Teklu2015, Schulze2018}, \textsc{Horizon-AGN} \citep{Dubois2016}, \textsc{Illustris-TNG100} \citep{Pillepich2018, Naiman2018, Springel2018, Nelson2019}, and the chemodynamical simulation of \cite{Taylor2015, Taylor2017}. 
For this work, the first in a series, we make use of the MAGPI theoretical dataset from {\sc Eagle}. In particular, we use the ready-to-use mock kinematic data cubes of the 481 {\sc Eagle} galaxies classified as MAGPI Primary Targets \citep{Foster2021}, created using \textsc{SimSpin v2.8.3} \footnote{\url{https://kateharborne.github.io/SimSpin/}}. Here we provide a summary of how these mock observations are created. The full description can be found in \cite{Harborne2020, Harborne2023}.

\textsc{SimSpin} acts as a virtual telescope wrapper, where the parameters of the virtual telescope can be chosen in a variety of different ways in order to reproduce observations of any IFS instrument. For the MAGPI theoretical dataset, the mock datacubes are created to match the MAGPI observations: a spatial pixel scale of 0.2 arcsec, a velocity pixel scale of 1.25 \AA, an average line-spread function of full-width half-maximum = 2.63 \AA, and observational ‘noise’. The galaxies are projected to a redshift distance of $z$ = 0.3 to match the observation specifications. To best reproduce the variety of inclination angles we see in observations, the mocks are generated at a random inclination angle.
The resulting data cubes provide values for the observed flux, mass, $v$, $\sigma$, as well as the higher-order kinematics moments $h_3$ and $h_4$ \citep{Harborne2023}. 

The MAGPI \textsc{Eagle} theoretical library also provides different measurements for the effective radius, $R_{\rm e}$. Here we take the photometric major axis as measured using the {\sc ProFound} {\sc getEllipses} function \citep{Robotham2018profound}, as our $R_{\rm e}$. 

\textsc{SimSpin} also gives the user the 3D shape of the galaxy at $R_{\rm e}$. The shapes are computed using the method described in the work of \cite{Bassett2019},  which in turn is based on the work from \cite{Li2018prolate} and \cite{Allgood2006}. A short summary of the method is presented here, and the full description can be found in \cite{Harborne2023}. See \cite{Zemp2011, Valenzuela2024} for an overview of different methods.

To start, the initial distribution of stellar particles is assumed to be an ellipsoid with axis ratios $p = q = 1$ (i.e. a sphere, where $p = b/a$ and $q = c/a$, with $a, b$ and $c$ representing the axis lengths in decreasing size such that $a \geq b \geq c$ and $p \geq q$, by necessity). This ellipsoid is grown from the centre position assigned until it contains half the total stellar mass within the subhalo. Stellar particles are then used within the region to measure the reduced inertia tensor. 
The eigenvalues and eigenvectors of this tensor, $I_{ij}$ give the orientation and distribution of matter within the ellipsoid. The ellipsoid is then deformed to match the distribution of stellar particles, and the procedure is repeated, with a new ellipsoid grown to contain half the mass that reflects the new matter distribution. This process is repeated until the axis ratios $p$ and $q$ stabilise over ten iterations.

We take these final $p$ and $q$ as our axis ratios at 1 $R_{\rm e}$ and we compute the triaxial parameter $T_{\rm Re}$, calculated at 1 $R_{\rm e}$ and defined in \cite{Binney2008}: 
\begin{equation}\label{eq:t_re}
    T_{\rm Re} = (1 - p_{\rm Re}^2)/(1 - q_{\rm Re}^2).
\end{equation}

\subsection{Particle data - Mass profiles} \label{section:mass}

The masses of {\sc Eagle} galaxies are measured using spherical apertures. This gives similar results to the 2D Petrosian aperture used in observational works, but provides an orientation-independent mass measurement for each galaxy.  Following previous studies based on the {\sc Eagle} simulations, we define the total stellar mass of a galaxy as the mass of the stellar particles within a 50~pkpc (pkpc refers to proper kpc) aperture centred at the centre of potential (e.g. \citealt{Schaye2015}). We note that selecting a different aperture (i.e. 30~pkpc) does not change our results. Similarly, dark matter (DM) masses and total masses are computed in spherical apertures (aperture values are publicly available and can be accessed from the {\sc Eagle} database \footnote{https://virgodb.dur.ac.uk/}). For each of our galaxies, we calculate the enclosed stellar, dark, and total mass within 1 $R_{\rm e}$ by interpolating the values between the two radial apertures closest to 1 $R_{\rm e}$.

We note that due to the resolution of the simulation, dynamical properties of galaxies are not reliable for stellar masses $\lesssim 10^{9.5}\,\rm M_{\odot}$ (see Appendix~A in \citealt{Lagos2017}, and \citealt{Ludlow2023}). Moreover, \cite{Ludlow2023} showed that a stellar mass $\gtrsim 10^{10}\,\rm M_{\odot}$ is needed for stellar kinematics to be resolved at the 3D half-mass radii of {\sc Eagle} galaxies; at lower masses, gravitational scattering between stellar and DM particles affects their kinematics and spatial distribution. It is important to note that DM mass resolution and gravitational scattering can also affect the circularity distributions of stellar disc particles by transforming galactic discs from flattened structures into rounder spheroidal systems, causing them to lose rotational support in the process \citep{Wilkinson2023}.

\subsection{Particle data - Orbital Distributions}\label{section:orbits}
We use the orbital distributions and decomposition of stellar particles in {\sc Eagle} galaxies calculated in \cite{Santucci2024}. Here, we provide a summary of the method.

To get to the orbital distribution and decomposition of stellar particles in {\sc Eagle} galaxies, we first define the $z$-axis of a galaxy as that parallel to the stellar angular momentum vector of the galaxy computed with all stellar particles of the subhalo with the spatial reference frame being centred on the position of the most bound particle, and the velocity reference being the mean mass-weighted velocity of all stellar particles in the subhalo.

We then calculate the ratio between the z-component of the specific angular momentum of a stellar particle relative to the specific angular momentum of a circular orbit of the same energy, $\lambda_{\rm z}$ (a.k.a. circularity parameter), following the method introduced by \cite{Abadi2003}.
With the calculation above, we construct a 2-dimensional distribution of $\lambda_{\rm z}$ vs projected distance (in the $\rm xy$ plane of the galaxy) for each galaxy.

We then separate orbits into four different components as follows: 
\begin{itemize}
    \item {\it cold orbits} selected as $\lambda_{\rm z} > 0.8$, are those with near-circular orbits;
    \item {\it warm orbits} selected as $0.25 < \lambda_{\rm z} < 0.8$, are somewhat disc-like, but with a considerable amount of angular momentum out of the disc ($z$) plane;
    \item {\it hot orbits} selected as $-0.25 < \lambda_{\rm z} < 0.25$, are those with near radial orbits;
    \item {\it counter-rotating orbits} are those with $\lambda_{\rm z} < -0.25$.
\end{itemize}
These thresholds were adopted to follow the analysis of \cite{Zhu2018nature, Jin2020, Santucci2022} and D24, and allow for easier comparison with observations.

However, these values of the circularity parameter, $\lambda_{\rm z}$, are instantaneous, while the model-derived circularities are time-averaged. This can create biases when comparing the circularity distributions from the models and the true ones. The stellar particles in our simulated galaxies do not necessarily conserve $\lambda_{\rm z}$ when orbiting in the potential, especially those particles on radial/box orbits. These orbits (as mentioned above) are usually characterised by $\lambda_{\rm z} \sim 0$. This means that in the instantaneous orbital distributions particles on radial/box orbits can appear to have the same orbital circularity as particles on cold/warm orbits (i.e. "spreading" the circularity distribution). A way to obtain the orbits’ circularity in a similar manner to the time-averaged obtained from the dynamical modelling is to freeze the potential, integrate the particle orbits in the potential, and calculate the average values along the orbits. This method, however, is computationally and time expensive. Here for simplicity, since we aim to compare the orbital distributions of simulated galaxies with those calculated in observations and mock-observations using Schwarzschild dynamical models, which are time-averaged, we calculate $\lambda_{\rm z}$ for each particle in three different snapshots (snapshot 24 - at $z=0.36$, snapshot 25 - at $z=0.27$, and snapshot 26 - at $z=0.18$) and we then average them to get an orbital distribution that is as close as possible to a time-averaged one. A similar approach has also been used by Reiter et al., in prep, for TNG50 galaxies, who found this method to be able to better reproduce time-averaged orbits. We note, however, that TNG50 has a high time-resolution between snapshots ($\sim$ 150 Myr), while the time range between two {\sc Eagle} snapshots is close to 1 Gyr. 

From the orbital distributions we calculate the fraction of orbits in each component, within 1 $R_{\rm e}$, by summing the mass of each stellar particle within 1~$R_{\rm e}$ that is in the corresponding orbital family, and dividing by the total stellar mass within 1~$R_{\rm e}$.


\subsection{{\sc Eagle} Sample Selection}\label{sec:selection}

We select our {\sc Eagle} galaxies to match the MAGPI primary targets. This initial selection gives us 481 galaxies with 10.0 $< \log M_{\star, 50 \rm kpc}/M_{\odot} <$ 11.54. We exclude 4 of these galaxies where $R_{\rm e} < 0.6$ arcsec, due to their spatial size being smaller than the mock instrumental spatial resolution. Since our focus is on the dynamical properties within 1 $R_{\rm e}$, we select galaxies with robust kinematic data (with at least 50 particles in each spaxel; \citealt{Jimenez2023}, Appendix and Harborne et al. in prep) available up to at least 1 $R_{\rm e}$. We carefully visually inspect the 350 selected galaxies and we exclude all galaxies whose kinematics are influenced by mergers, or that have a bright secondary object within 1 \re~in their stellar velocity field. This leaves us with 259 galaxies with 10.12 $< \log M_{\star, 50 \rm kpc}/M_{\odot} <$ 11.54. After further data quality requirements imposed in the below sections, we have a final sample of \Nsample~galaxies.

\section{Schwarzschild models - {\sc{DYNAMITE}}}
The MAGPI theoretical dataset was created with the intention to be used as a like-to-like comparison for MAGPI observational data. As such, we strive to apply methods as closely as possible to those used for the MAGPI data in order to reduce any possible bias due to different methods.

The two basic requirements when constructing a Schwarzschild model are 2D stellar kinematics with as many higher-order moments as feasible (dependent on the quality of the spectral data), and a description of the galaxy's surface brightness. In the following sections we describe our modelling of the galaxy surface brightness and our extraction of the 2D stellar kinematics.
Additionally, since models for the MAGPI sample were obtained by transforming the surface brightness models to stellar mass models via spectral stellar population measurements to infer stellar mass-to-light ratios, we also choose to use stellar mass maps instead of luminosity maps to construct our mass distribution.
\subsection{Data preparation}
\subsubsection{Multi Gaussian Expansion profiles}
We parameterise the galaxy mass using Multi Gaussian Expansion (MGE) method, which consists of a series of concentric 2D Gaussians \citep{Emsellem1994}, using the Python package \textsc{mgefit} \citep{Cappellari2002}. This method enables us to take the PSF into account; given a value of the inclination and assuming an intrinsic shape, the MGE model can be deprojected \emph{analytically}, and used to describe the intrinsic mass density of the stellar tracer within a given gravitational potential. 

\begin{figure}
\centering
\includegraphics[width=\linewidth]{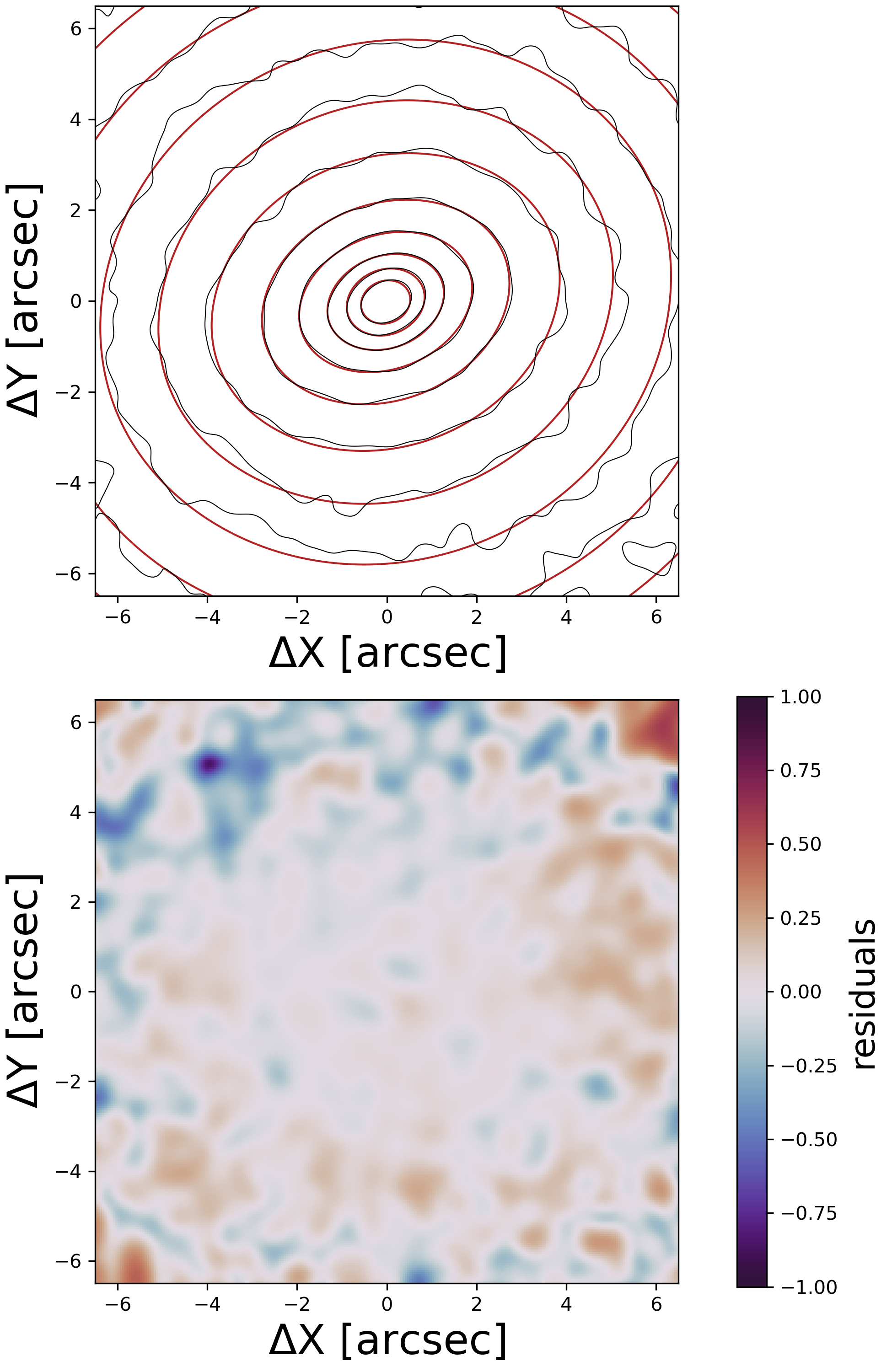}
\caption{The stellar mass model for example galaxy {\sc Eagle} 9498015. The top panel shows the isophotes of the galaxy in black and of the best MGE model overlaid in red. The bottom panel shows the residuals of the model as (data-model) divided by data image.}
\label{fig:mge}
\end{figure}

We fit the MGEs by following the process described in \cite{Derkenne2023} and D24 as closely as possible. Similarly to them, we require all components to have the same position angle, as a change in position angle was not warranted by the fit quality of the models compared to the data. The fits are applied to the mass images created by \textsc{SimSpin}. We convert the dispersion of each Gaussian component from units of pixels to arcseconds by multiplying by the mock-MUSE pixel scale (0.2 arcsec). We show an example of an MGE fit in Figure~\ref{fig:mge}. 

\begin{figure}
\centering
\includegraphics[width=\linewidth]{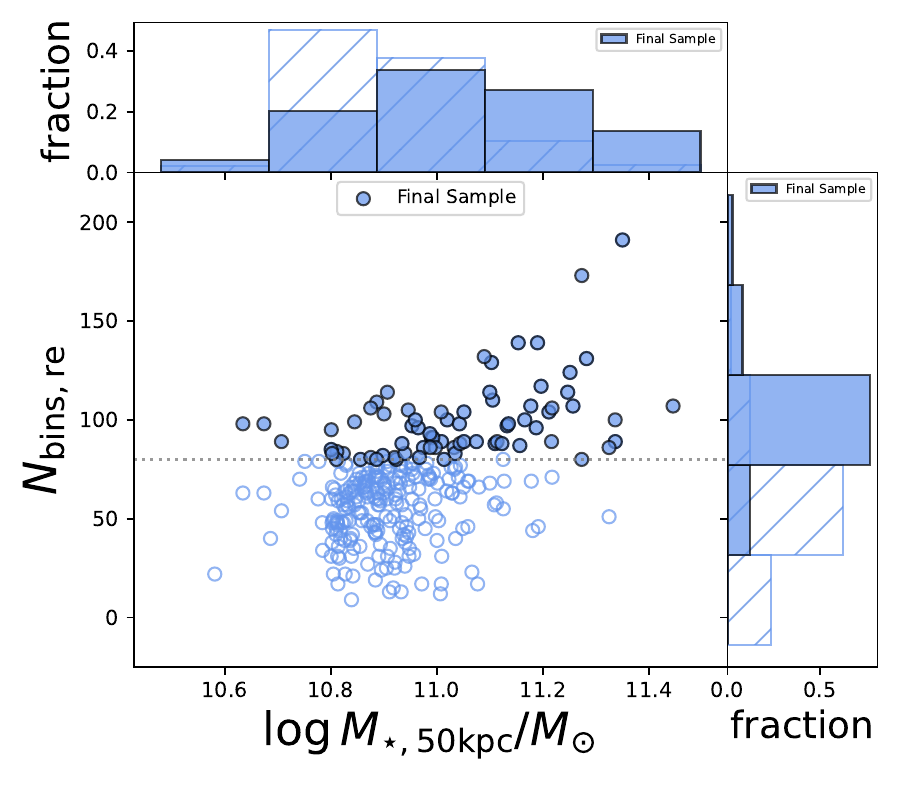}
\caption{Number of Voronoi bins within 1 $R_{\rm e}$ for galaxies that meet our quality criteria as a function of stellar mass within 50 kpc for our {\sc Eagle} sample (Sec. \ref{sec:selection}). We exclude from our sample galaxies that have fewer than 80 Voronoi bins within 1 $R_{\rm e}$ (horizontal grey dashed line). Galaxies that do not meet the required number of Voronoi bins within 1\re~ are shown as open circles. The histograms on the top and on the right panels show the distribution in mass and Number of Voronoi bins, respectively, for the galaxies in our sample before (hatched histograms) and after (coloured histograms) the cut in Voronoi bins. The average number of Voronoi bins for each galaxy in the final sample is 99. }
\label{fig:bins_mass}
\end{figure}
\subsubsection{Stellar Kinematics}
As mentioned in Sec. \ref{sec:simspin}, stellar kinematic maps are one of the outputs from \textsc{SimSpin} provided for the MAGPI theoretical dataset. These include line of sight velocity ($v$), velocity dispersion ($\sigma$), and  $h_{\rm 3}$ and $h_{\rm 4}$, the Gauss-Hermite moments which describe the skewness and kurtosis of the profile, respectively \citep{Gerhard1993, vanderMarel1993}. 

\begin{figure*} 
\centering
\includegraphics[width=\linewidth]{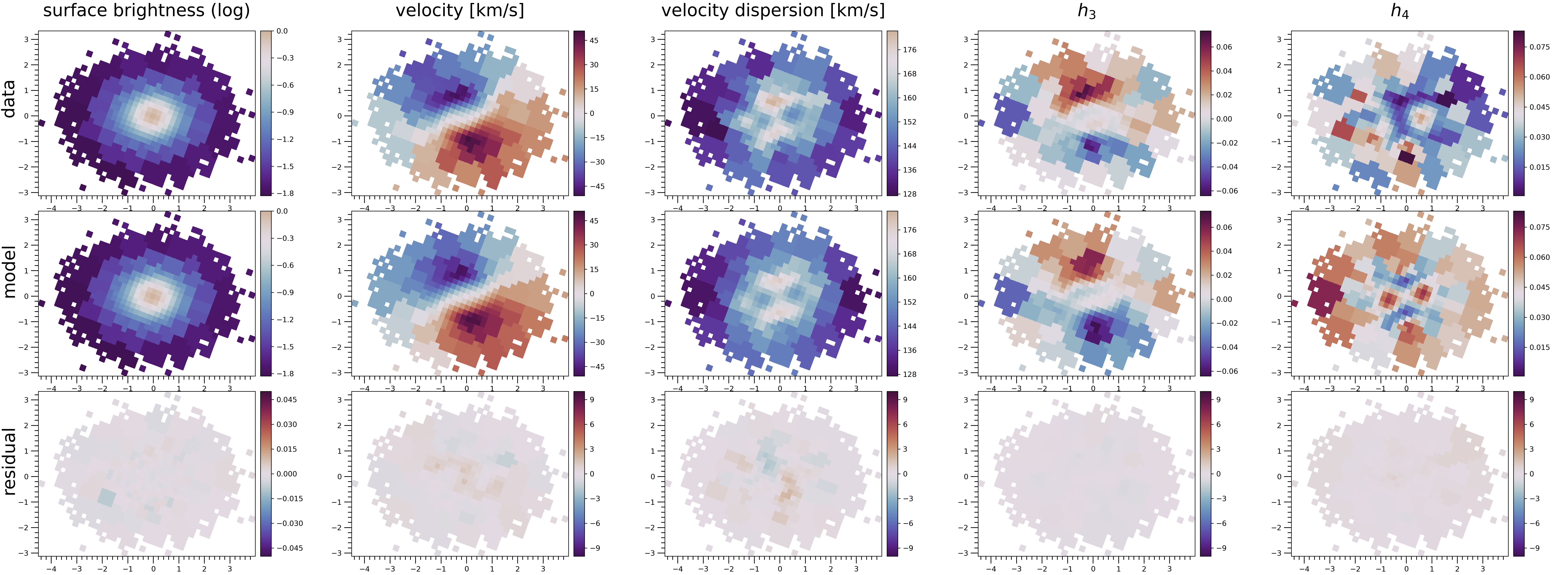}
\caption{Example galaxy {\sc Eagle} 9498015. This galaxy has $\log M_{\star, 50 \rm kpc}/M_{\odot} = 11.19$ and $R_{\rm e} = 6.08$ kpc. Columns show 2D maps for, from left to right, flux, velocity, velocity dispersion, $h_3$ and $h_4$. The first row shows the observed maps, the second row shows the best-fit maps derived from the Schwarzschild modelling and the third row shows the residuals, calculated as the difference between the observation and the model, divided by the observational uncertainties (or divided by the model, in the case of the surface brightness). The best-fit model maps ($\chi^2_{\rm red} = 0.72$) accurately reconstruct the structures seen in the observations. }
\label{fig:bf_example}
\end{figure*}

\begin{figure*} 
\centering
\includegraphics[width=\linewidth]{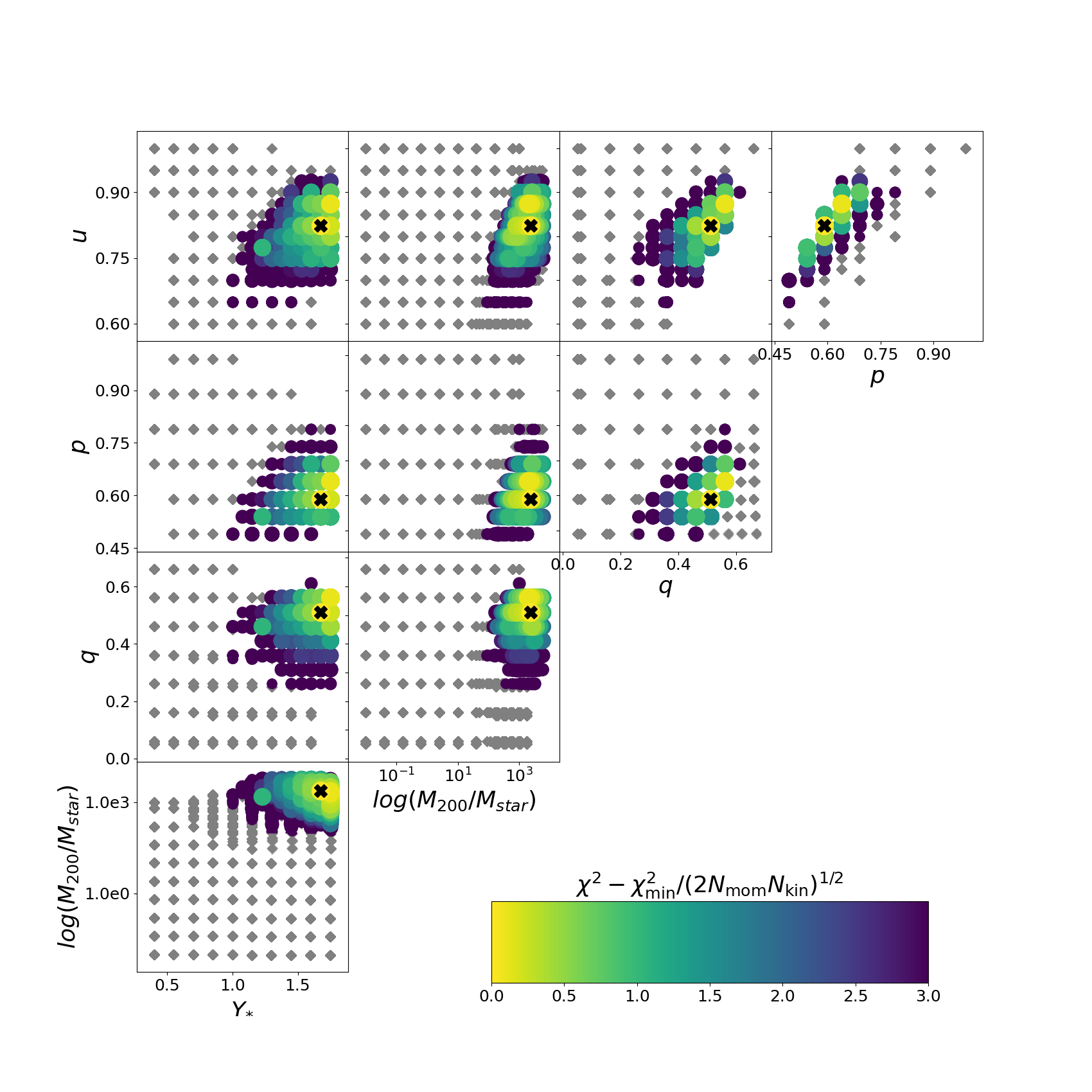}
\caption{Example galaxy {\sc Eagle} 9498015: model parameter grid. There are five free parameters: the intrinsic shape of the flattest Gaussian component ($p_{\rm min}$, $q_{\rm min}$, $u_{\rm min}$), the DM-to-stellar mass ratio, $\log M_{200}/M_{\star}$, and a global mass scaling factor, which scales both stellar and dark components jointly, $\Upsilon_{*}$. The diamonds represent the parameters explored, with the best-fit model highlighted with a black cross. Models within the best-fit region are colour-coded according to their $\chi^2$ values shown in the colour bar. The best-fit values are well-constrained. All 8336 models are shown, which is a typical number of models run for each {\sc Eagle} object.}
\label{fig:param_space}
\end{figure*}
We apply Voronoi binning to the kinematic maps - using the number of particles ($N_{\rm p}$) as a proxy for signal-to-noise. We select a target $N_{\rm p} = 200$ and apply the Voronoi binning with the Python package {\sc{vorbin}} \citep{Cappellari2003}. We find that this threshold is the best compromise to achieve robust values for the kinematic maps (ensuring the outer regions, where the uncertainties are higher and the kinematic less reliable, have less weight) while keeping a high enough number of bins for our galaxies to be properly spatially sampled (as demonstrated by Harborne et al., in prep). For each Voronoi bin, $v$, $\sigma$, $h_{\rm 3}$, and $h_{\rm 4}$ are taken as the weighted average (by number of particles) of the values in each spaxel belonging to the Voronoi bin. Spaxels with more than 200 particles are left unbinned. We show the number of bins within $R_{\rm e}$ as a function of stellar mass in Fig. \ref{fig:bins_mass}. To ensure we have enough resolution elements in our galaxies to construct robust models, we further exclude all the galaxies with less than 80 Voronoi bins within 1 $R_{\rm e}$ \footnote{We find that the properties derived from models built using maps with fewer than 80 bins are not well constrained, even when thousands of models are calculated.}. This cut leaves us with \Nsample~galaxies. It is important to note that this final selection excludes most of the smallest galaxies from our sample, as well as low-mass edge-on (with angles as defined in Sec. \ref{sec:simspin}) galaxies, since they have fewer bins within 1~\re~ than more face-on galaxies of similar size and mass.

\subsection{Construction of Schwarzschild models} \label{sec:schwarz}
Schwarzschild models \citep{Schwarzschild1979} are a flexible but computationally expensive method that allows us to model triaxial stellar systems in three steps: firstly we construct a model for the underlying gravitational potential of the galaxy; secondly we calculate a representative library of orbits in these gravitational potentials; and then we find a combination of orbits that can reproduce the observed kinematic maps and luminosity distribution. These steps are fully described in \cite{vandenBosch2008} and \cite{Zhu2018spirals}. Here we use the triaxial {\sc{DYNAMITE}} code\footnote{version: 3.0.0} \citep{DYN2020} to construct the Schwarzschild models, with orbit mirroring corrections included \citep{Quenneville2021b,Thater2022dyna}. The models are computationally expensive as a large library of orbits in a given gravitational potential are integrated for a set number of orbital periods (to ensure reliable convergence), and the resulting surface brightness and kinematic moments are compared to observations. Any change to the given potential therefore results in re-integrating the orbit library to determine the orbital weights that best reproduce the observations via non-negative least squares (NNLS) optimisation. This can require running thousands of models to characterise a parameter space, and large orbit libraries in order to encapsulate the true diversity of orbital structures that may be present in galaxies. For the construction of our models, we set-up the gravitational potential, orbit libraries, and parameter space searches in a similar way to D24. We present an outline of our configuration process below.

The underlying gravitational potential of the galaxy is constructed with a triaxial stellar component, a spherical dark matter halo, and a central super-massive black hole. The triaxial stellar component mass is calculated from the best-fit two-dimensional MGE. Each of these MGEs mass densities is de-projected assuming the orientation in space of the galaxy, described by three viewing angles ($\theta$, $\phi$, $\psi$), to obtain a three-dimensional mass density. The space orientation  ($\theta$, $\phi$, $\psi$) can be converted directly to the intrinsic shape, described by the parameters $p=B/A$, $q=C/A$, $u=A'/A$, where $A$, $B$, and $C$ represent the major, medium, and minor axes of the 3D triaxial Gaussian, respectively, and $A'$ is the projected major axis of the 3D triaxial Gaussian. We leave $p$, $q$, $u$ as free parameters to allow intrinsic triaxial shapes to be fitted. 

A spherical Navarro-Frenk-White (NFW; \citealt{Navarro1996}) halo is adopted. The mass, $M_{200}$ (mass enclosed within a radius, $R_{200}$, where the average density is 200 times the critical density), in an NFW dark matter halo is determined by two parameters in our modelling: the concentration parameter, $c$, and the ratio of dark matter within $R_{200}$, $M_{200}/ M_{\star}$ (where $M_{\star}$ is the total stellar mass). Due to the MAGPI and MAGPI mocks data quality, we adopt a $M_{200}-c$ coupled halo. This fixes the concentration according to the relation measured by \cite{Dutton2014}, leaving the DM-to-stellar mass ratio within $R_{200}$, $M_{200}/ M_{\star}$, free. 

For completeness, we also include a central super-massive black hole component. However, the spatial resolution of our data cannot resolve the influence radius of the black hole, so its mass leaves no imprint on the stellar kinematic maps of our galaxies and therefore does not affect our results. For this reason, similarly to D24, we include a black hole with fixed mass, derived by the redshift dependent mass-velocity dispersion relation from \cite{Robertson2006}. The total (luminous plus dark) mass of the system can then be scaled by the global constant ($\Upsilon_{*}$), as the scaled model can be evaluated by scaling the orbital velocities without re-integrating the orbit library.

Our Schwarzschild implementation creates initial conditions for the orbits by sampling from the three integrals of motion (energy $E$, second integral $I_2$ and third integral $I_3$). Each set of orbits is sampled across the three integrals with the following number of points: $n_E \times n_{I_2} \times n_{I_3} = 21 \times 15 \times 9$. We test different samplings of the second integral $I_2$, for example, increasing $n_{I_2}$ from 15 to 18 and to 40, and re-fitting the models. The best-fit values retrieved by this search are consistent, within the 1$\sigma$ confidence level, with the best-fit values retrieved by our regular runs. We use three sets of $21 \times 15 \times 9$ orbits: a set of box orbits, tube-orbits, and counter-rotating tube-orbits. Following D24, we also do not dither the orbits, since dithering is not necessary for MAGPI-like data (see Appendix A of D24), nor do we regularise\footnote{\cite{Lipka2021} presented a novel technique to optimise the amount of regularisation used for orbit-weight solving, which can have a significant impact on constraining physical parameters such as galaxy mass, anisotropy and viewing angle. We did not implement this technique due to the data being of poorer resolution than the one employed in \cite{Lipka2021}, and hence adding regularisation would not yield more precise results in our case} the orbital weights.

Once the orbit libraries are created, the orbit weights are determined by fitting the set of orbit-superposition models to the projected and de-projected luminosity density (from MGE fits) and the two-dimensional line-of-sight stellar kinematics. The modelled and the observed kinematics are then divided by the observational error so that a $\chi^2$ comparison is achieved. The weights are determined by the \citet{vandenBosch2008} implementation, using the \citet{Lawson1974} non-negative least squares implementation. The best-fit model is defined as the model with minimum kinematic $\chi^2$.To ensure that we get the best deprojection fit, we first run what we call a `coarse' grid of models across the allowed parameter space for the five free parameters, sampling a minimum of 3 evenly spaced values within the bounds for $p$, $q$, and $u$, at least 6 values in $\Upsilon_{*}$, and at least 7 values in DM-to-stellar mass ratio $M_{200}/ M_{\star}$ between -2 and 3 in log space. 

The best-fit model from the coarse run is then used to initiate a dynamic `fine' parameter search with increasingly smaller step sizes around the best-fit solution, up to a minimum of $0.05$ in log space for dark matter fraction, 0.02 for $p$ and $q$, 0.01 for $u$  (which generally tends very close to unity), and 0.1 in $\Upsilon_{*}$. The coarse and fine model runs use the `FullGrid' and `LegacyGridSearch' options, respectively,  in the {\sc{DYNAMITE}} code.

As in \cite{Zhu2018spirals} and D24, we define the $1\sigma$ confidence level as $\Delta \chi^2 = \sqrt{2N_{\mathrm{kin}}N_{\mathrm{mom}}}$, where $N_{\mathrm{kin}}$ is the number of kinematic Voronoi bins and $N_{\mathrm{mom}}$ is the number of kinematic Gauss-Hermite moments. We choose this definition to ease the comparison with previous work, and particularly with D24. The quoted parameter value (e.g. $p$) is calculated from the model with the lowest $\chi^2$ value. To estimate the uncertainties, parameter values from all $1\sigma$ models are calculated, and the upper and lower bounds of that distribution are taken. For example, the upper error is quoted as $\mathrm{max}(\textbf{X}) - x_0$, where $\textbf{X}$ is the vector of all $1\sigma$ model parameter values and $x_0$ is the value from the lowest $\chi^2$ model. We show the best-fit kinematic maps and parameter space coloured by Schwarzschild model $\chi^2$ value for example galaxy {\sc Eagle} 9498015 in Fig. ~\ref{fig:bf_example} and ~\ref{fig:param_space}.

From these models we derive the internal orbital structure, inner mass distribution, intrinsic shape, and velocity anisotropy for each galaxy. Intrinsic shapes can also be reconverted into viewing angles. The viewing angles are typically well-constrained, with a median uncertainty of 7 degrees. In addition, following \citet{Zhu2018nature, Jin2020, Santucci2022, Santucci2024}, we separate orbits into four different components (cold, warm, hot, and counter-rotating) according to their orbit circularity $\lambda_z$ as described in Sec. \ref{section:orbits},
with $\lambda_z \equiv J_z/J_{\rm max}(E)$ around the short $z$-axis, normalized by the maximum of a circular orbit with the same binding energy $E$. We calculate the fraction of orbits in each component within 1 $R_{\rm e}$. 

We derive dynamically-based intrinsic shapes using the triaxial parameter at 1 $R_{\rm e}$, $T_{\rm Re} = (1 - p_{\rm Re}^2)/(1 - q_{\rm Re}^2)$, where $p_{\rm Re}$ and $q_{\rm Re}$ are the intermediate-to-long and short-to-long axis ratios at 1 \re. We separate galaxies into three groups according to this parameter: oblate ($T_{\rm Re} < 0.3$), triaxial ($0.3 \leq T_{\rm Re} < 0.8$) and prolate ($T_{\rm Re} \geq 0.8$).

A detailed analysis of these properties, and how they compare to the particle-derived properties is presented in the following section.

\section{Results}
In this section, we present the results we obtain modelling our sample of \Nsample~ MAGPI-like {\sc Eagle} galaxies with the Schwarzschild modelling technique. First, we compare the properties derived from the models with the true values from the simulations in Sec. \ref{sec:onetoone}. We then present analysis of the intrinsic shapes and orbital structures of the sample, comparing our results with results from the MAGPI Survey (D24) in Sec. \ref{sec:orbs_analysis}. 

\subsection{Comparison with particle data} \label{sec:onetoone}

\subsubsection{Mass profiles}
\begin{figure} 
\centering
\includegraphics[width=\linewidth]{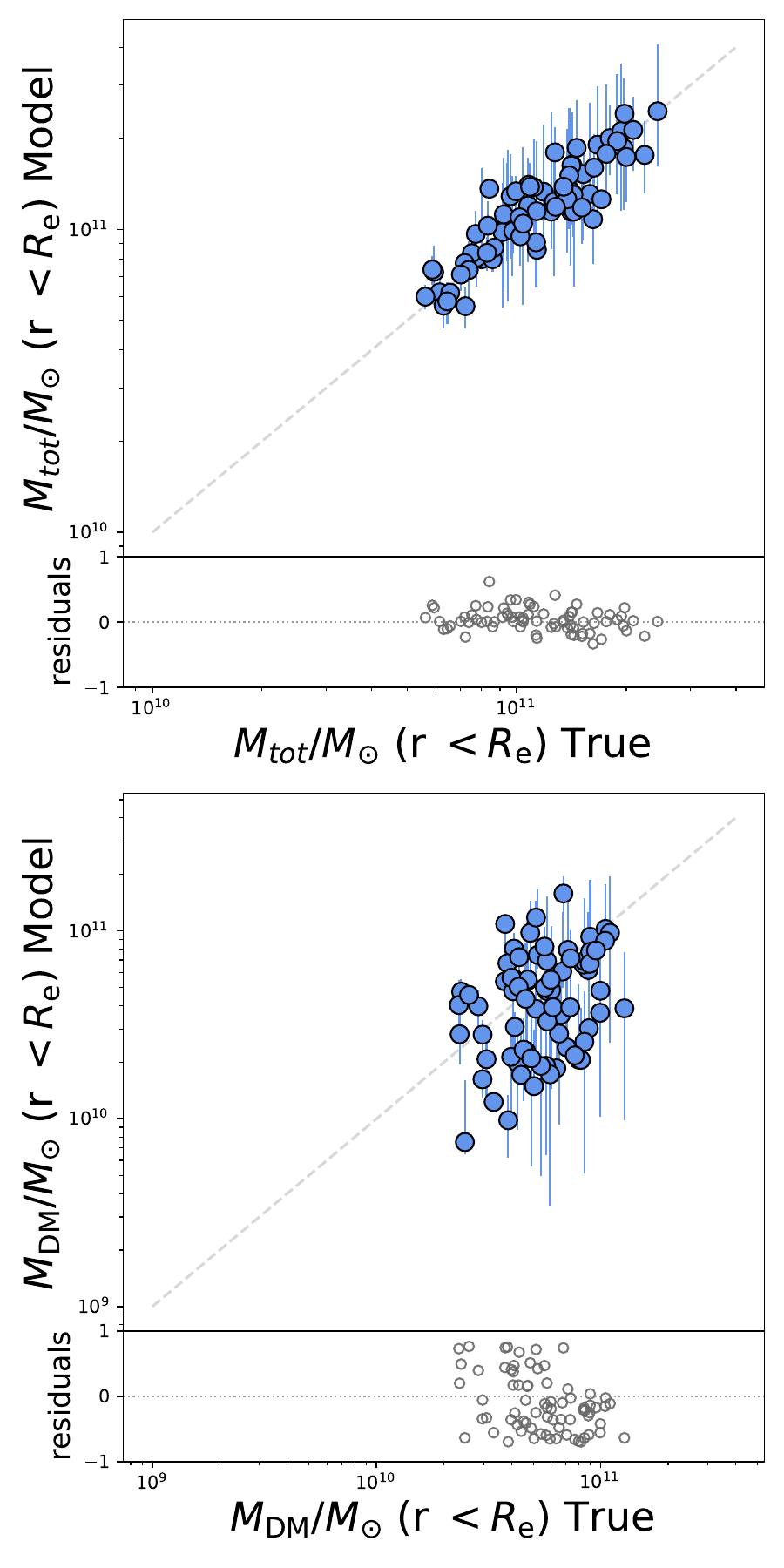}
\caption{One-to-one comparison of the true and model-recovered total (top panel) and dark (bottom panel) mass of {\sc Eagle} galaxies within 1 $R_{\rm e}$. Fractional residuals (defined as (model-true)/true) of the total and dark matter mass, respectively, are shown in the bottom region of each panel. The one-to-one line is shown in grey. The model-reconstructed enclosed total mass well matches the true one, within the errors. The derived enclosed DM mass is in general in agreement with the true one, however there are several galaxies where the DM mass is underestimated.}
\label{fig:eagle_mass}
\end{figure}

The total mass ($M_{\rm tot}$) radial distribution is one of the fundamental parameters of the Schwarzschild model, which includes a stellar component and a dark matter component. As mentioned in Sec. \ref{sec:schwarz}, a black hole mass component is included as well, but not discussed here as its contribution to the total mass distribution is negligible. Fig. \ref{fig:eagle_mass} shows the one-to-one comparison of the model-recovered and the true value of the enclosed (within 1 $R_{\rm e}$) total mass (top panel), and dark matter mass (bottom panel). 

\begin{figure} 
\centering
\includegraphics[width=\linewidth]{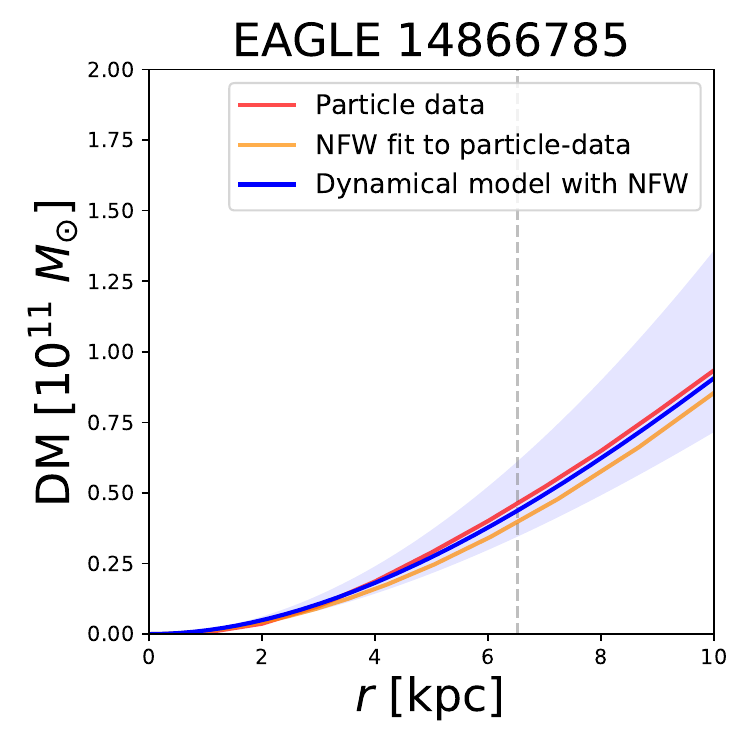}
\includegraphics[width=\linewidth]{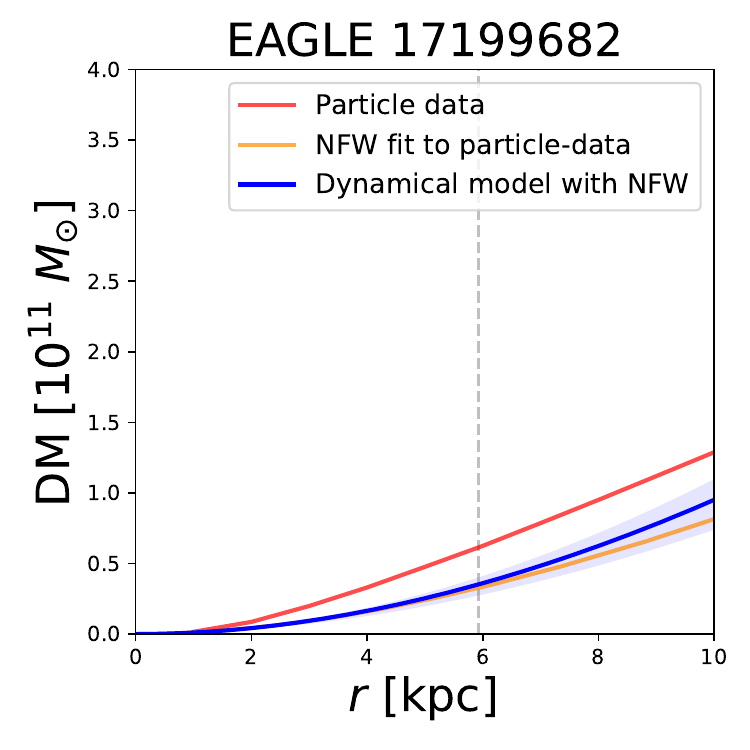}
\caption{Enclosed DM mass profiles from the particle data (in red), the NFW fit on that data (in dark orange, from \citealt{Schaller2015}), and the enclosed DM mass retrieved by our modelling (shaded area delimits the 1$\sigma$ region). The top panel shows the DM mass profiles for a galaxy with a DM distribution close to the NFW one, while the bottom panel shows the enclosed DM mass for a galaxy where the NFW profile is not a good fit. As seen, even though our fit does not reproduce the real DM distribution, it is consistent with the NFW fit on that galaxy.}
\label{fig:dm_profiles}
\end{figure}

In general, the total mass is recovered well, with a median offset of 9\%. However, we find that for almost half of the galaxies in our sample the retrieved value of DM  within 1 $R_{\rm e}$ has a median offset of 48\%, in the majority of cases underestimating the true value. While there are degeneracies in the models that could lead to a not well-constrained DM profile, we find that this discrepancy is mostly due to the assumed NFW profile not being able to reproduce the DM distribution in the galaxy - likely because the DM distribution is not spherical, for example being more concentrated than an NFW. Moreover, for lower-mass galaxies, the models are constrained up to about 1 \re, which is not enough to well constrain the dark matter content.

\cite{Schaller2015} investigated the mass distributions of {\sc Eagle} central galaxies, fitting their halo densities with NFW profiles, focusing on the outer regions of the haloes. We compare the enclosed DM mass from their NFW fits with our results, for the galaxies in both samples. In Fig. \ref{fig:dm_profiles} we show the enclosed DM mass profiles from the particle data, the NFW fit on that data (from \citealt{Schaller2015}), and the enclosed DM mass retrieved by our modelling. The top panel shows the DM mass profiles for a galaxy with a DM distribution close to the NFW one, while the bottom panel shows the enclosed DM mass for a galaxy where the NFW profile is not a good fit in the inner region (although it is a good fit in the outer region of the halo, which it is not shown here). As seen, even though our fit does not reproduce the real DM distribution, it is consistent with the NFW fit on that galaxy. The retrieved value of DM mass within 1 \re~ is therefore affected by the fact that the assumed NFW profile is not able to reproduce the DM distribution in the inner regions of the galaxy. \cite{Schaller2015} also showed that while the {\sc Eagle} haloes are well described by the NFW density profile at large radii, closer to the centre the presence of stars can produce steeper profiles. We believe these differences in the inner density profiles, as well as the possible non-spherical shape of the DM haloes, can lead to the discrepancies we see in the enclosed DM mass profiles. 

\subsubsection{Intrinsic shapes}
\begin{figure} 
\centering
\includegraphics[width=\linewidth]{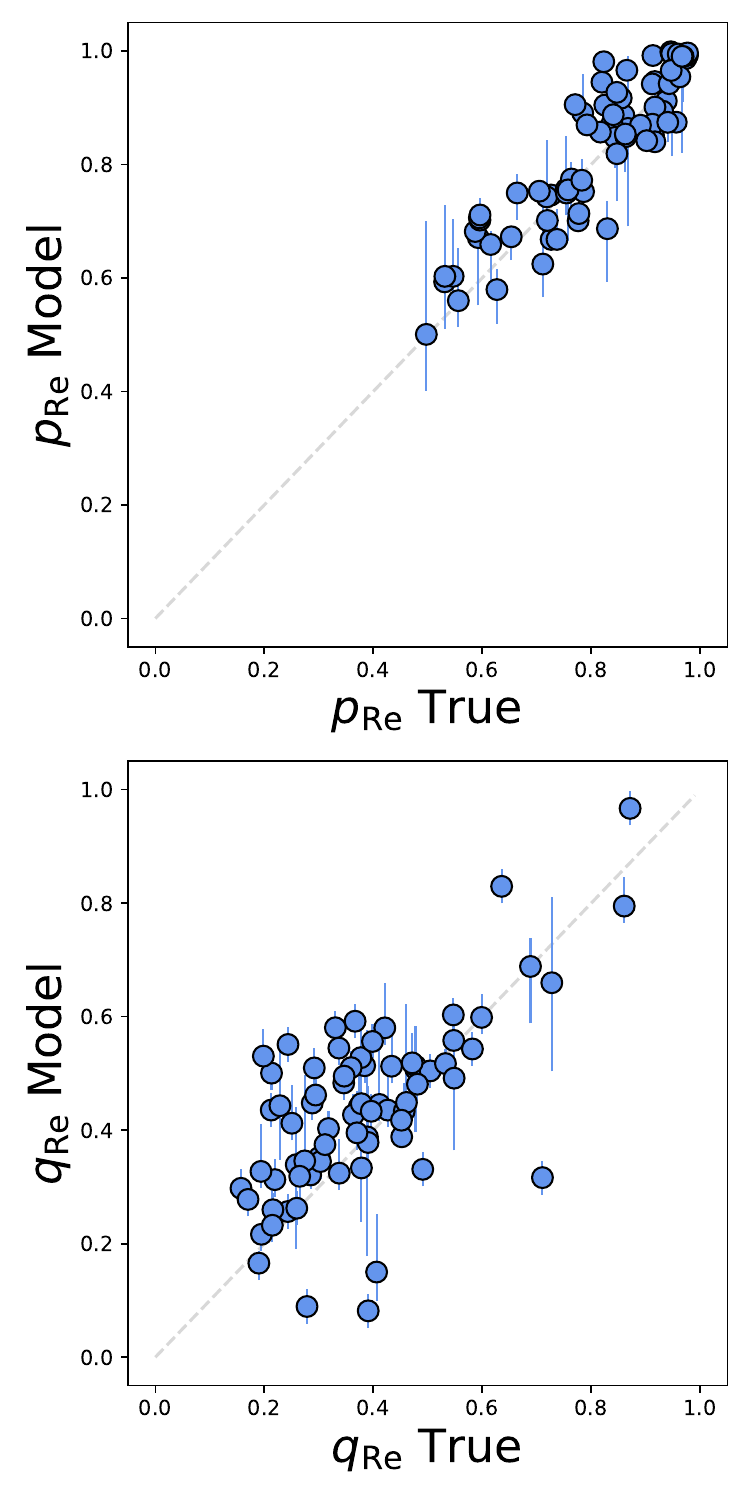}
\caption{One-to-one comparison of the true and model-recovered axis ratios within 1\re, $p_{\rm Re}$ (top panel) and $q_{\rm Re}$ (bottom panel). We find that the model-derived $p_{\rm Re}$ is consistent, within the errors, with the true value. $q_{\rm Re}$ shows a large scatter around the one-to-one line, with a median overestimation of 0.18. }
\label{fig:pq_re}
\end{figure}
We show the one-to-one comparison between the model-derived and the true axis ratios within 1\re, $p_{\rm Re}$ and $q_{\rm Re}$, in Fig. \ref{fig:pq_re}. We find that the model-derived $p_{\rm Re}$ is consistent, within the errors, with the true value, with a median overestimation of 0.04. The model-derived $q_{\rm Re}$ is qualitatively consistent with the true value, but shows a large scatter around the one-to-one line, with a median overestimation of 0.18.

\begin{figure} 
\centering
\includegraphics[width=\linewidth]{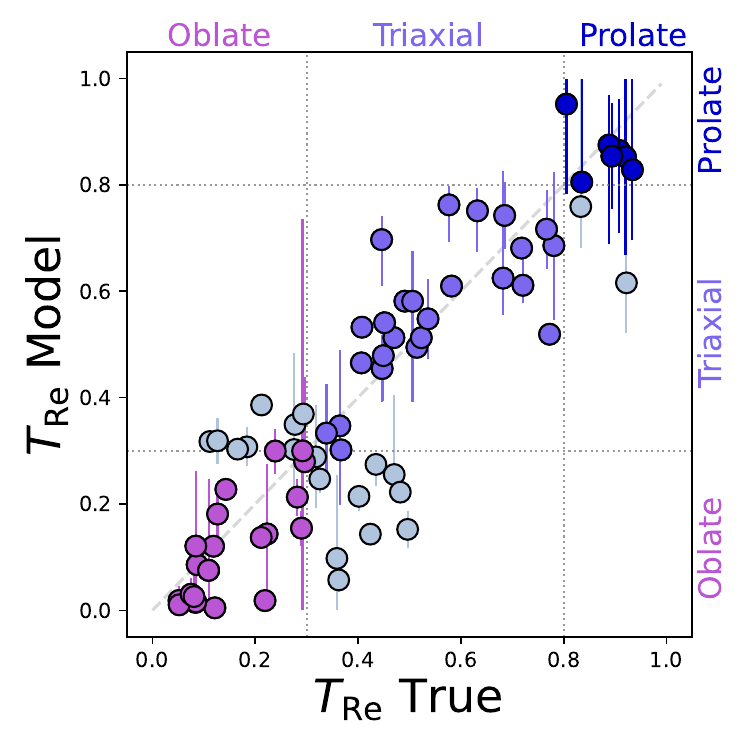}
\caption{One-to-one comparison of the true and model-recovered triaxial parameter $T_{\rm Re} = (1 - p_{\rm Re}^2)/(1 - q_{\rm Re}^2)$. Galaxies with $T_{\rm Re} < 0.3$ are classified as oblate, galaxies with $T_{\rm Re} \geq 0.8$ as prolate and those in-between as triaxial. Grey dashed lines represent $T_{\rm Re} = 0.1$, $T_{\rm Re} = 0.3$ and $T_{\rm Re} = 0.8$. Galaxies that are classified as oblate using both their true and model-derived triaxiality are shown in orchid, galaxies classified as triaxial are shown in purple and galaxies classified as prolate are shown in blue. Galaxies that are misclassified using the model-derived triaxiality are shown in grey. In general, we find that, even when the values of triaxiality are not close to the one-to-one line, 73\% of the galaxies in our sample are correctly classified using their model-derived triaxiality. }
\label{fig:t_re}
\end{figure}

Fig. \ref{fig:t_re} shows the one-to-one comparison of the true and model-recovered triaxiality at 1 $R_{\rm e}$ of our galaxies. We find that the true value of triaxiality is recovered with a median offset of 0.07 (underestimating the true values). However, there is large scatter around the one-to-one line, mostly at the low and at the high end of the triaxiality parameter. Similar results were also found by \cite{Jin2019}, looking at nine galaxies in the highest resolution simulation, Illustris-1, of the Illustris simulations. It is interesting to note that even though we find less agreement between the model-derived and the true $q_{\rm Re}$, it is $p_{\rm Re}$ that has the largest impact on $T_{\rm Re}$, so that even a small difference ($\sim$ 0.02, for example) between the model-derived and the true $p_{\rm Re}$ can lead to very different values of $T_{\rm Re}$ (by construction, see Eq. \ref{eq:t_re}). Moreover, the triaxiality parameter and the axis ratios are closely connected with the inclination angle of the observed galaxy - which can sometimes be difficult to constrain, in particular for galaxies where there is no clear disc structure nor triaxial features. 

In order to understand the impact of the discrepancy between modelled and true shape, we separate galaxies into three groups according to their triaxiality: oblate ($T_{\rm Re} < 0.3$), prolate ($T_{\rm Re} \geq 0.8$) and triaxial (0.3 $\leq T_{\rm Re} < 0.8$). We find that 73\% of the galaxies in our sample that are classified as oblate, triaxial or prolate using the true values, maintain the same classification when using the model-derived values. This shows that even though the exact values of individual galaxies could be off, in studying a bulk population we are able to get a statistical view of their shapes. 8 of the true oblate galaxies (11\% of the total and 26\% of the true oblate galaxies) are classified as triaxial, 10 (14\%) of the true triaxial galaxies are classified as oblate, and 2 of the true prolate galaxies (3\% of the total) are classified as triaxial. In the majority of cases, the reason behind the wrong classification of the 18 oblate and triaxial misclassified galaxies is the low ($<30^{\circ}$) inclination angle, which can be difficult to constrain in the modelling. 
    
\subsubsection{Fraction of orbits}

\begin{figure*} 
\centering
\includegraphics[width=\linewidth]{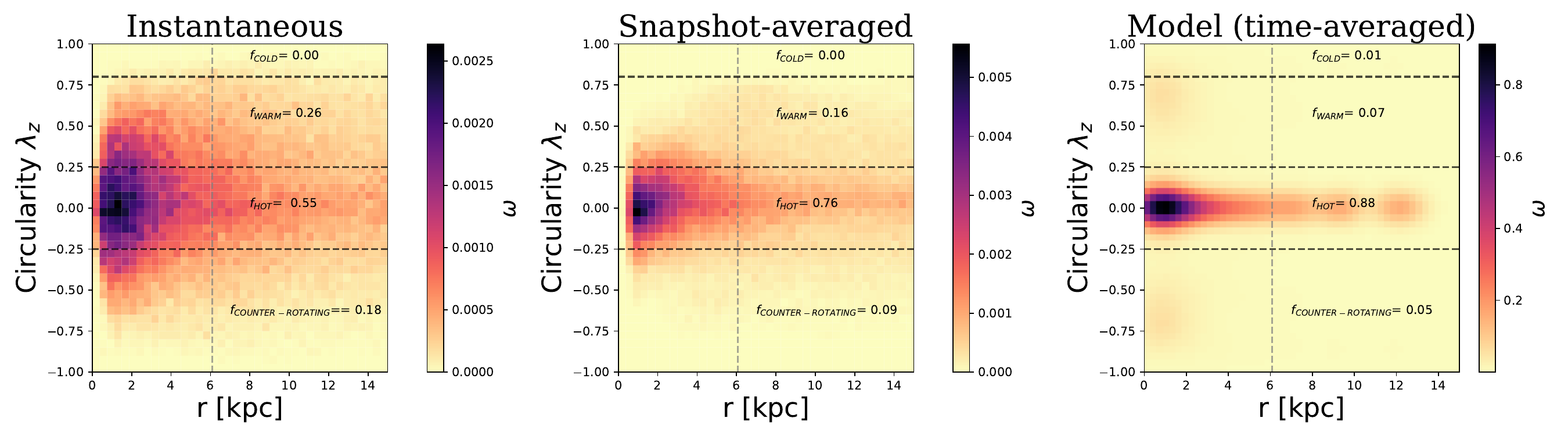}
\caption{Example galaxy {\sc Eagle} 9498015: orbit circularity $\lambda_z$ versus radius (in kpc). The colour indicates the density, $w$, of the orbits on the phase space, and the three horizontal black dashed lines indicate $\lambda_z$ = 0.8, $\lambda_z$ = 0.25 and $\lambda_z$ = -0.25, dividing the orbits into four regions (cold, warm, hot and counter-rotating orbits). Fractions of mass within 1\re~in each component are annotated in the figures. The vertical grey dashed line is located at 1$ R_{\rm e}$. Left panel shows the true instantaneous orbital distribution of the stellar particles; middle panel show the average orbital distribution calculated over 3 snapshots; right panel shows the time-averaged orbital distribution derived from the dynamical modelling. Averaging over the 3 snapshots helps to reconstruct the orbital distribution and makes it possible to compare the particles' circularity with what is derived by {\sc{DYNAMITE}}. }
\label{fig:circularity}
\end{figure*}

As mentioned in Sec. \ref{section:orbits}, stellar orbits can be characterized by two main properties: the radius $r$ (in this case time-averaged), representing the size of each orbit, and their circularity $\lambda_z$. In the Schwarzschild models, these orbital distributions are time-averaged over 200 orbital periods. Taking the radius, $r$, and the circularity, $\lambda_z$, of each orbit, and considering their weights given by the solution from the best-fit model, we can use the orbit circularity distribution in the phase space to obtain the probability density of orbits within 1 $R_{\rm e}$, for each galaxy. An example of the circularity space can be seen in Fig.~\ref{fig:circularity} for example galaxy {\sc Eagle} 9498015, which is dominated by hot orbits. The panel on the left shows the particles' instantaneous orbital distribution, while the middle panel shows the snapshot-averaged one and the left panel shows the orbital distribution from the models (time-averaged over 200 orbital periods). The fractions of mass in each orbital component within 1 \re~ are annotated in each panel. These figures show how the difficulty is not just in recovering the orbital distributions, but also in finding the best way to compare them between simulations and models, since the techniques used to derive them are different. Instantaneous orbital distributions show stellar particles covering almost the entire range in $\lambda_z$. As mentioned in Sec. \ref{section:orbits}, the stellar particles in our simulated galaxies do not necessarily conserve $\lambda_{\rm z}$ when orbiting in the potential. This is particularly true for particles on radial/box orbits (which have, averaging over time, $\lambda_{\rm z} \sim 0$). Instead, some of these particles will appear to have the same orbital circularity as particles on cold/warm orbits (i.e. "spreading" the circularity distribution). This difference between the instantaneous (from the particle data) and the time-averaged (from the models) orbital distributions means that we cannot directly compare them (see Appendix \ref{app:frozen_pot} for more details). For this reason, we use the snapshot-averaged orbital distributions.

\begin{figure*} 
\centering
\includegraphics[width=\linewidth]{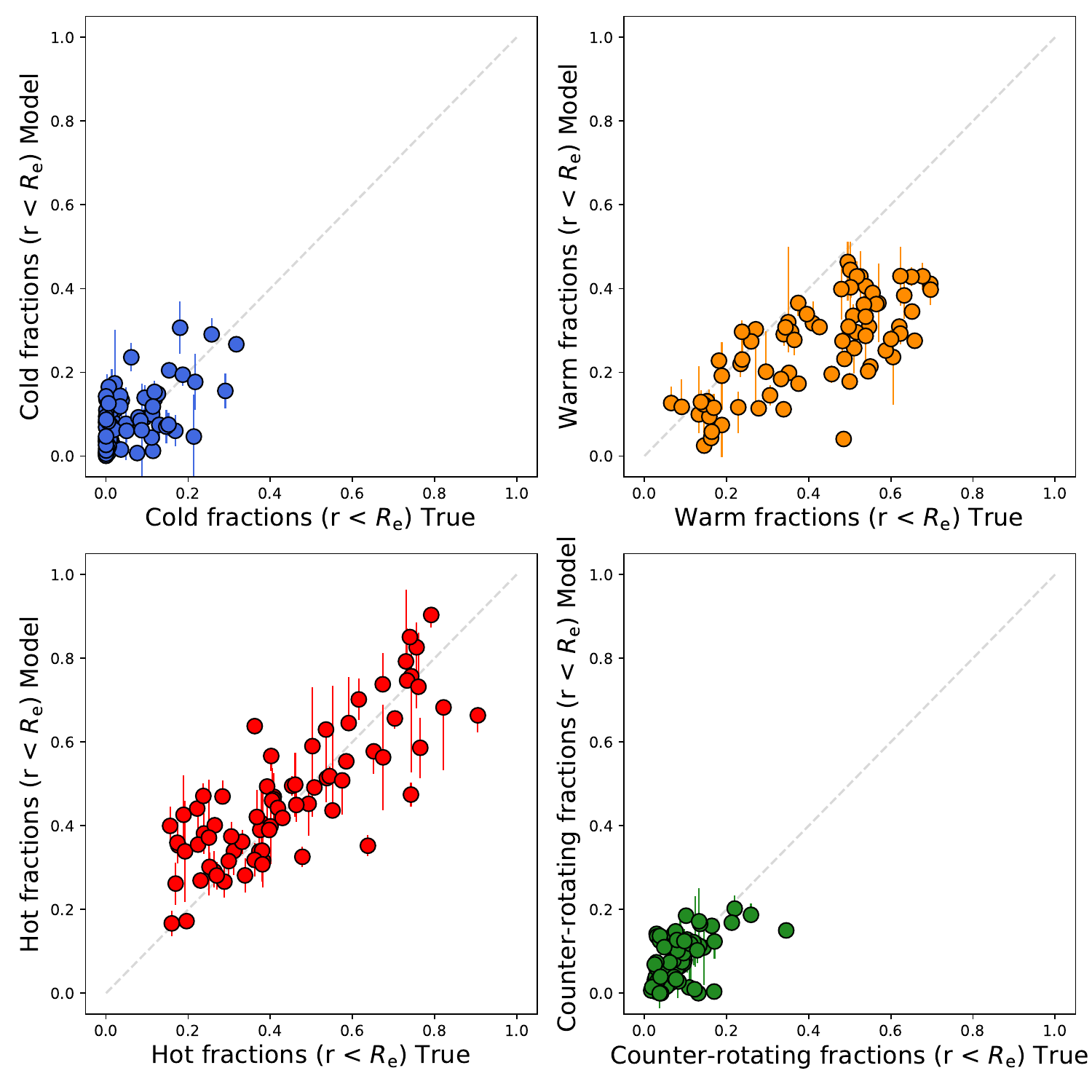}
\caption{One-to-one comparison of the true and model-recovered fraction of the cold (in blue), warm (in orange), hot (in red), and counter-rotating (in green) orbits. The true values are calculated from the average of the orbital distribution in three different snapshots. We find that, for $\lambda_z$, the comparison between the model-derived and true fractions of orbits are generally in agreement, within the errors. However, fractions of hot orbits tend to be overestimated and fractions of warm orbits tend to be underestimated by the models.}
\label{fig:orbits}
\end{figure*}

From the cuts defined in Sec.~\ref{sec:schwarz}, we calculate the orbital fractions for the sample within $R_{\rm e}$ and then compare them to their true values averaged over 3 snapshots. Fig. \ref{fig:orbits} (as described in Sec. \ref{section:orbits}) shows the one-to-one comparison of the true and model-recovered fraction of the cold, warm, and hot orbits.
We find that, for $\lambda_z$, the comparison between the model-derived and true fractions of cold and counter-rotating orbits are in general agreement, within the errors. The model-derived fractions of hot orbits are higher (especially at low fractions) than those derived from the particle data, while the warm fractions are generally lower (especially at high fractions). 
We find that the model-derived fractions of cold orbits have a median overestimation of 0.07, while the warm fractions are underestimated by 0.14, and the counter-rotating fractions of 0.02. The hot orbit fractions have a median overestimation of 0.06. We stress, however, that the discrepancies that we see are most likely not caused by the Schwarzschild models not being able to reproduce the orbital distributions well, but by the two different methods used to derive the orbits' circularity.

We also show the one-to-one comparison of the instantaneous and model-recovered fractions of the cold, warm, hot and counter-rotating orbits in Appendix \ref{app:orbits}, Fig. \ref{fig:orbits_inst}, for completeness. We find that the instantaneous orbits and the model-derived orbits have large scatter and offsets, compared to the snapshot-averaged values. In particular, as expected, the model-derived fractions of hot orbits are generally higher than the instantaneous fractions, while the model-derived fractions of warm orbits tend to be lower than the instantaneous fractions. As mentioned in Sec. \ref{section:orbits}, this is not surprising since the model-derived orbital distributions are time-averaged.

\subsection{Dark matter content, intrinsic shapes, and orbital distributions of massive galaxies in {\sc Eagle} and MAGPI} \label{sec:orbs_analysis}

We now explore the distribution of the model-derived galaxy properties (DM fractions, intrinsic shapes, fraction of orbits, and velocity anisotropy) with stellar mass and as a function of other galaxy properties, and we compare our results with those derived for 22 MAGPI galaxies by D24.
\begin{figure} 
\centering
\includegraphics[width=\linewidth]{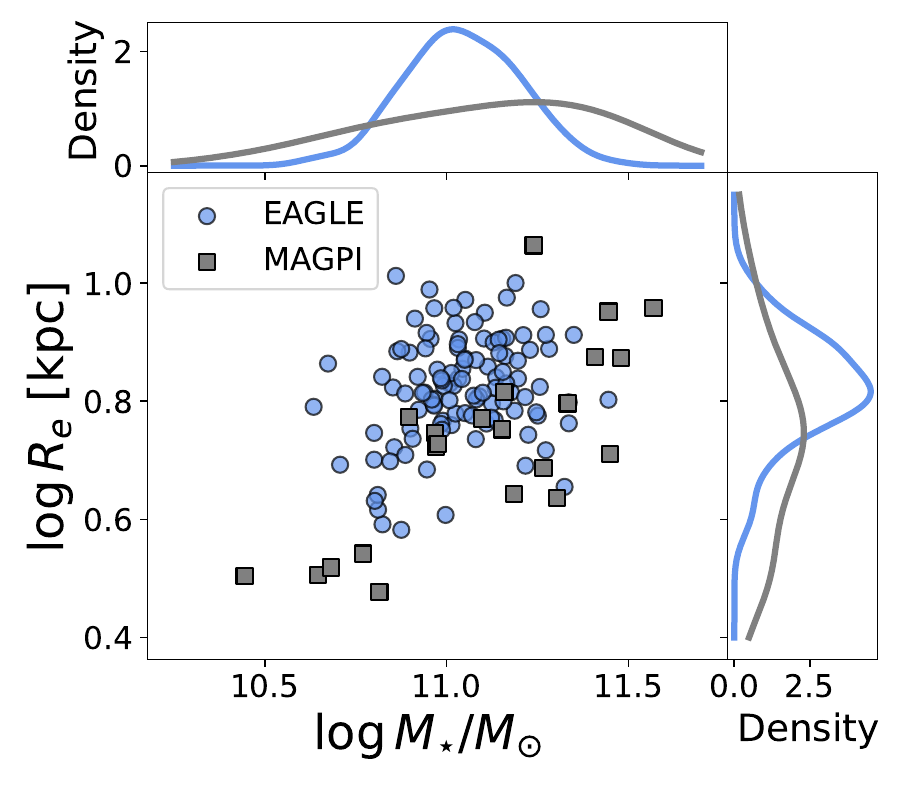}
\caption{Distribution of the {\sc Eagle} galaxies in our sample (light blue circles) and of the MAGPI galaxies in D24 (grey squares) in the size-stellar mass plane. We show the marginalised distributions (calculated using kernel density estimates) of mass and size of the galaxies in each sample in the top and right panels of the figure, respectively. Light blue lines are for the {\sc Eagle} galaxies in our sample, while grey lines are for the galaxies in the MAGPI sample. {\sc Eagle} galaxies are larger than the galaxies in the MAGPI sample. The two samples also differ in stellar mass: our {\sc Eagle} sample has relatively more galaxies in the stellar mass range $10.8 < \log M_{\star, 50 \rm kpc}/M{\odot}< 11.2$ than the MAGPI sample. }
\label{fig:eagle_magpi_distr}
\end{figure}

We show in Fig. \ref{fig:eagle_magpi_distr} the distributions in the size-mass plane for our sample of {\sc Eagle} galaxies, in light blue, and the MAGPI galaxies in D24, in grey. {\sc Eagle} galaxies are larger than the galaxies in the MAGPI sample. This is likely a result of our Voronoi binning selection criterion, which excluded the most compact galaxies from our sample. The two samples also differ in stellar mass: our {\sc Eagle} sample has relatively more galaxies in the stellar mass range $10.8 < \log M_{\star, 50 \rm kpc}/M{\odot}< 11.2$ than the MAGPI sample. 

\begin{figure} 
\centering
\includegraphics[width=\linewidth]{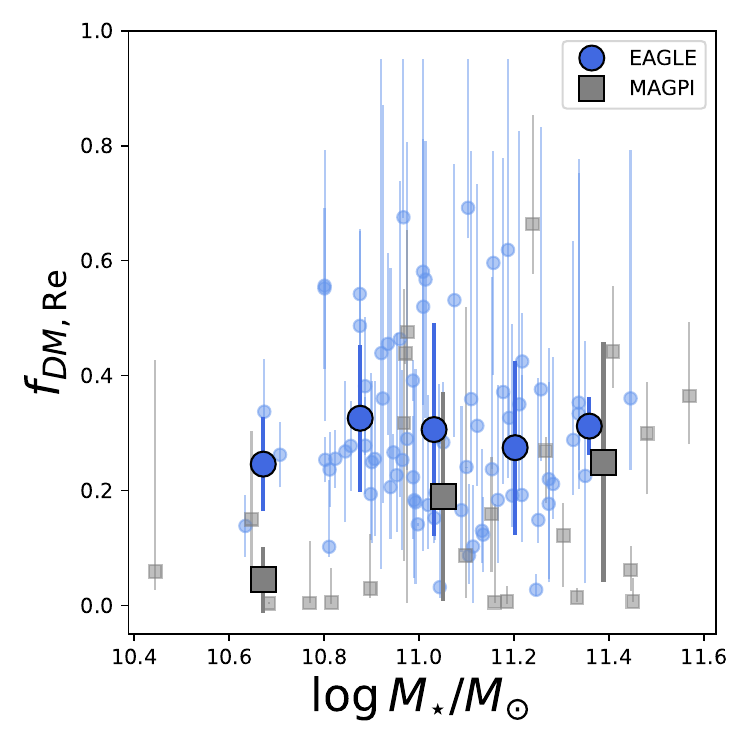}
\caption{Fraction of dark matter to total mass ($f_{\rm{DM}} = M_{\rm dark}/M_{\rm tot}$) within 1 $R_{\rm e}$ from dynamical models as a function of stellar mass for {\sc Eagle} galaxies (light blue circles) and MAGPI galaxies (grey squares). Bold points show the mean values for each mass bin, and the errobars show their 1$\sigma$ scatter. We find a median DM fraction of 26\%, with a standard deviation of 15\%. {\sc Eagle} galaxies with stellar masses below $\log (M_{\star, 50 \rm kpc}/M_{\odot}) \sim 11.2$ show higher DM fractions than MAGPI galaxies, while {\sc Eagle} galaxies at higher stellar masses have values of DM fractions comparable to those of MAGPI galaxies.  }
\label{fig:dm_eagle_magpi}
\end{figure}

We define the fraction of DM within 1 $R_{\rm e}$, $f_{\rm DM}$, as the ratio between the enclosed mass of DM within 1 $R_{\rm e}$ and the enclosed total mass within 1 $R_{\rm e}$. We show in Fig. \ref{fig:dm_eagle_magpi} the model-derived DM fractions for our \Nsample~ {\sc Eagle} galaxies (in light blue) and for the 22 MAGPI galaxies (in grey) modelled by D24. To better compare the results, we also show the mean values of $f_{\rm DM}$ for different bins in mass. We find that the median DM fraction within the 1 $R_{\rm e}$ is 26\%, with a standard deviation of 15\% for the {\sc Eagle} sample. This fraction is larger, although still consistent within the errors, than what is found for MAGPI galaxies ($f_{\rm DM}$ = 10\% with a standard deviation of 19\%). This discrepancy is mainly driven by the large fractions of DM found for {\sc Eagle} galaxies with $\log (M_{\star, 50 \rm kpc}/M_{\odot}) < 11.2$. For galaxies with stellar masses above $\log (M_{\star, 50 \rm kpc}/M_{\odot}) \sim 11.2$, we find comparable DM fractions for {\sc Eagle} and MAGPI galaxies of similar stellar masses ($f_{\rm DM}$ = 26\% with a standard deviation of 9\% and $f_{\rm DM}$ = 27\% with a standard deviation of 21\%, respectively). We note, however, that the uncertainties in these fractions are large, and they are not taken in consideration when calculating the median values of DM in the different mass bins in Fig. \ref{fig:dm_eagle_magpi}. 

\begin{figure} 
\centering
\includegraphics[width=\linewidth]{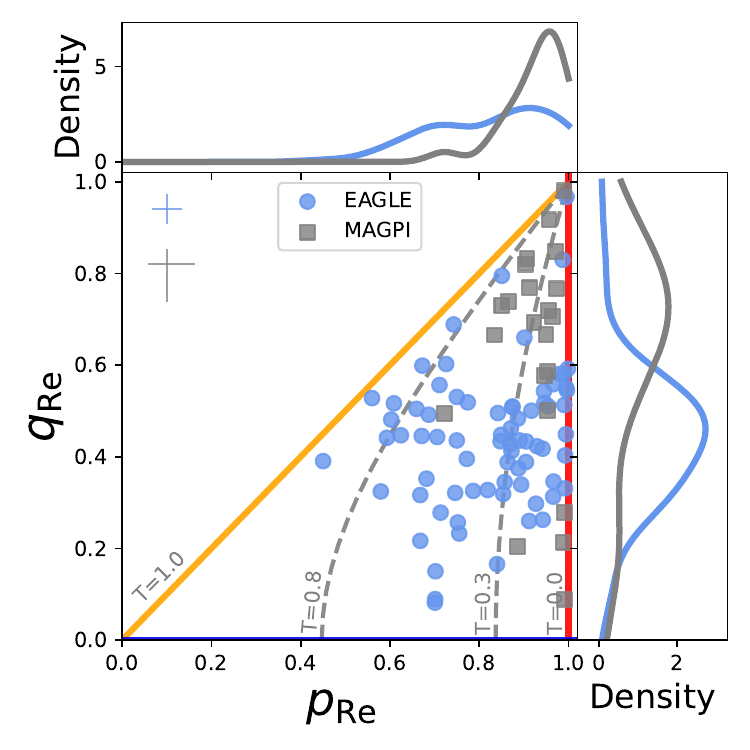}
\caption{The distribution of axis ratios $q_{\rm Re}$ versus $p_{\rm Re}$ at 1 $R_{\rm e}$ from the best-fit models for all galaxies in our {\sc Eagle} sample (light blue circles) and in the MAGPI sample (grey squares; D24). The grey dashed curves from right to left indicate $T_{\rm Re} = 0.3$, and  $T_{\rm Re} = 07$; the orange line indicates $T_{\rm Re} = 1.0$, and the red line is for $T_{\rm Re} = 0.0$. Perfectly oblate spheroids have zero triaxiality. The blue and grey daggers in the top left corner show the median (asymmetric) errors on the intrinsic axial ratios for {\sc Eagle} and MAGPI, respectively. We show the marginalised distributions (calculated using kernel density estimates) of $q_{\rm Re}$ and $p_{\rm Re}$ of the galaxies in each sample in the top and right panels of the figure, respectively. Light blue lines are for the {\sc Eagle} galaxies in our sample, while grey lines are for the galaxies in the MAGPI sample. The majority of the galaxies in both the {\sc Eagle} sample and the MAGPI sample have non-oblate shapes.}
\label{fig:needle_disk_sphere}
\end{figure}

In Fig. \ref{fig:needle_disk_sphere} we present the intrinsic shapes as the distribution of axis ratios $q_{\rm Re}$ versus $p_{\rm Re}$ for all galaxies in the {\sc Eagle} (light blue circles) and MAGPI (grey squares) samples. The grey dashed curves separate the galaxies into oblate ($T_{\rm Re} < 0.3$), triaxial ($0.3 \leq T_{\rm Re} < 0.8$) and prolate ($T_{\rm Re} \geq 0.8$) galaxies. We find that the majority of galaxies are consistent with being ellipsoids of varying degrees of triaxiality, with a few tending towards circular (i.e. highly flattened, oblate) disks in their intrinsic shapes. Only 13/\Nsample \ {\sc Eagle} galaxies are classically defined oblate spheroids, with a triaxiality parameter of less than 0.1. The rest of the galaxies in the sample are oblate (19/\Nsample) and triaxial (35/\Nsample), with 7 prolate objects. MAGPI galaxies show different distributions in both $p_{\rm Re}$ and $q_{\rm Re}$, with values generally higher, in particular for $q_{\rm Re}$, than {\sc EAGLE} galaxies. Matching the samples in both mass range and size does not decrease the differences. However, we find that, despite these differences, MAGPI galaxies show a similar range in triaxiality as the {\sc Eagle} sample. The two samples have similar percentages of oblate spheroids: 18\% $\pm$ 4\% for {\sc Eagle} and 14\% $\pm$ 7\% for MAGPI. The MAGPI sample has a higher fraction (77\% $\pm$ 9\%)  of non-oblate ($T_{\rm Re}\geq 0.3$) galaxies than {\sc Eagle} (57\% $\pm$ 6\%). 

\begin{figure*} 
\centering
\includegraphics[width=\linewidth]{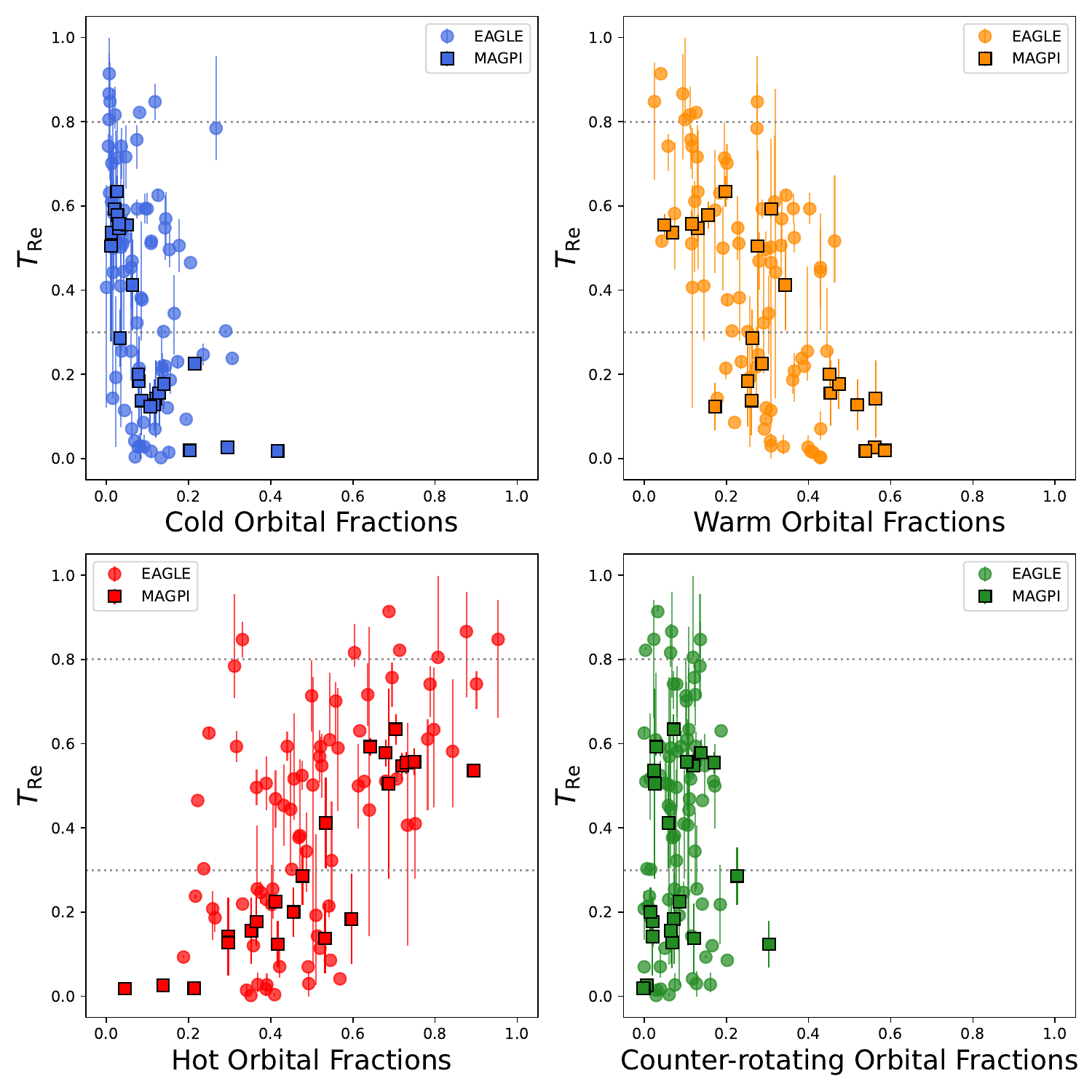}
\caption{Triaxiality, $T_{\rm Re}$ as a function of the fraction of orbits within 1 $R_{\rm e}$ for both {\sc Eagle} (coloured circles) and MAGPI (grey squares) galaxies. We find that the triaxiality parameter increases with increasing fraction of hot orbits and decreasing fraction of cold and warm orbits. No clear correlation is found between the fraction of counter-rotating orbits and a galaxy's triaxiality.}
\label{fig:fracs_te}
\end{figure*}

We also check whether we find a similar close connection for {\sc Eagle} galaxies between triaxiality and the fraction of hot orbits as found for MAGPI galaxies (D24). We show the triaxiality as a function of the fraction of orbits within 1 $R_{\rm e}$ for both {\sc Eagle} and MAGPI galaxies in Fig. \ref{fig:fracs_te}. We find that galaxies with higher triaxiality tend to have higher fractions of hot orbits and lower fractions of warm and cold orbits, while we do not see any trend with the counter-rotating orbits. To test whether these trends are statistically significant, we use the Kendall's correlation coefficient $\tau$, using the Python package \textit{scipy.stats.kendalltau} \citep{Virtanen2019}. This correlation coefficient is robust to small sample sizes. A $\tau$ value close to 1 indicates strong correlation, whereas a value close to $-$1 indicates strong anti-correlation. For the fractions of hot orbits, we find a value of $\tau = 0.36$, with a $p$-value of $4.29 \times 10^{-6}$. Negative correlations are found for the fraction of cold and warm orbits, with $\tau = -0.28$ and $p$-value of $4.57 \times 10^{-4}$ and $\tau = -0.41$ and $p$-value of $2.16 \times 10^{-7}$, respectively.


D24 found in their MAGPI sample that radially anisotropic ($\beta_r > 0$) objects are likely to be intrinsically triaxial and have mainly hot (box or radial) orbits, with few cold (short-axis tube) orbits. To check whether this is also true in our simulation sample, we define the velocity anisotropy parameter $\beta_r$, in the standard spherical coordinates ($r,\theta,\phi$), following \cite{Binney2008}: 
\begin{equation}
\beta_r = 1 - \frac{\Pi_{tt}}{ \Pi_{rr}},
\end{equation}
with 
\begin{equation}
\Pi_{tt} = \frac{\Pi_{\theta\theta}+\Pi_{\phi\phi}}{2},
\end{equation}

\begin{equation}\label{eq:D_kk}
\Pi_{kk} =  \int \rho \sigma_k^2 \,d^3x  = \sum_{n=1}^{N} M_n \sigma_{k,n}^2 
\end{equation}
with $\sigma_k$ the velocity dispersion along the direction $k$ at a given location inside the galaxy. The summation defines how we computed this quantity from our Schwarzschild models. $M_n$ is the mass contained in each of the $N$ polar grid cells in the meridional plane of the model, and $\sigma_{k,n}$ is the corresponding mean velocity dispersion along the direction $k$.
We calculate the value of $\beta_r$ at 1 $R_{\rm e}$, and we denote it $\beta_{\rm Re}$. $\beta_{\rm Re} > 0$ indicates radial anisotropy, $\beta_{\rm Re} < 0$ indicates tangential anisotropy and $\beta_{\rm Re} = 0$ indicates isotropy.

\begin{figure} 
\centering
\includegraphics[width=\linewidth]{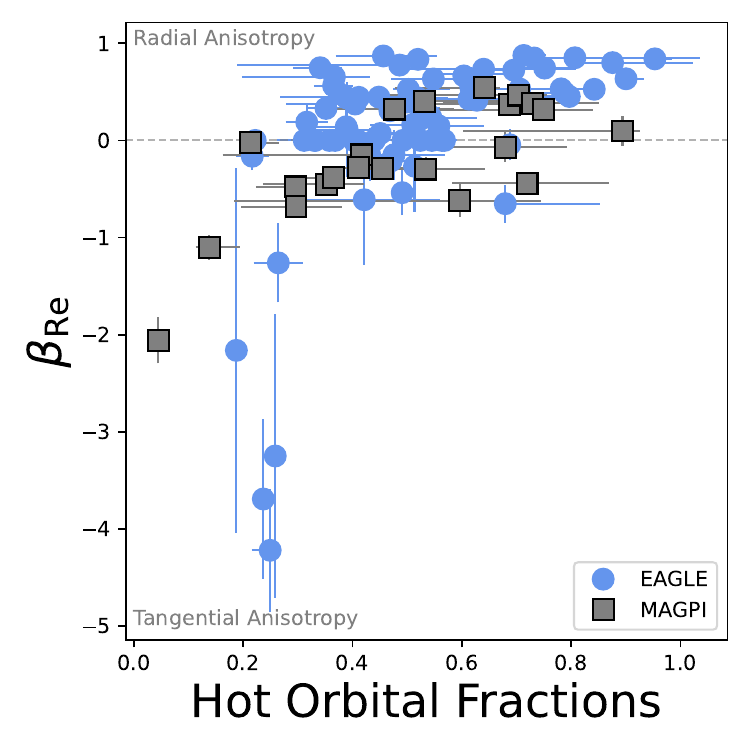}
\caption{Velocity dispersion anisotropy in spherical coordinates within 1 $R_{\rm e}$, $\beta_{\rm Re}$, as a function of the fraction of hot orbits for {\sc Eagle} galaxies (light blue circles) and MAGPI galaxies (grey squares). $\beta_{\rm Re} > 0$ indicates radial anisotropy, $\beta_{\rm Re} < 0$ indicates tangential anisotropy and $\beta_{\rm Re} = 0$ indicates isotropy. We find that $\beta_{\rm Re}$ increases with increasing fraction of hot orbits, so that galaxies with higher fraction of hot orbits are more radially anisotropic. We also find a good agreement between the MAGPI and {\sc Eagle} samples.}
\label{fig:f_hot_br}
\end{figure}

We show the velocity anisotropy $\beta_{\rm Re}$ as a function of the fraction of hot orbits, for {\sc Eagle} (light blue circles) and MAGPI (grey squares) galaxies in Fig. \ref{fig:f_hot_br}. We find a correlation between the fraction of hot orbits and $\beta_{\rm Re}$ with $\tau = 0.38$, and $p$-value $=2.04 \times 10^{-6}$, so that galaxies with higher fractions of hot orbits are more likely to be close to isotropy ($\beta_{\rm Re} = 0$) or radially anisotropic ($\beta_{\rm Re}> 0$), similar to what was found for MAGPI galaxies by D24 ($\tau = 0.47$, and $p$-value $=1.71 \times 10^{-3}$).

\section{Discussion}
In general, we find that Schwarzschild models of mock observations from {\sc Eagle} can, in most cases, reconstruct the true values of each parameter quite accurately, within the errors. However, there are parameters where the retrieved values are offset. One of these is the enclosed DM mass. We find that, while there are degeneracies in the models for which the enclosed DM mass can be harder to constrain, the majority of the discrepancy can be explained by the fact that we have assumed a spherical NFW distribution in the Schwarzschild modelling which, in some cases, it is not representative of the DM distribution in the simulated galaxies. We find a median underestimation of the enclosed DM mass of 48\%. It is interesting to note that previous results for the Illustris simulation \citep{Jin2019} have found that the model-derived dark matter within 1 $R_{\rm e}$ was generally overestimated (as opposed to the underestimation we find), while they found the stellar mass to be underestimated. This difference is most likely due to the fact that we are using stellar mass maps to construct our MGEs, instead of having light-based MGEs (as in \citealt{Jin2019}), helping to constrain the stellar mass.
Moreover, \cite{Valenzuela2024} found that the shape of the DM haloes in the Magneticum hydrodynamical cosmological simulation and found that, in the inner region of galaxies, the DM follows the stellar component in shape and orientation. Since {\sc DYNAMITE} allows for different types of DM distributions to be assumed, it would be interesting to test the systematics each assumption causes and their effect on the enclosed mass distributions in future work.

We also find that the derived intrinsic shapes within 1 $R_{\rm e}$ have large uncertainties, especially for galaxies with face-on viewing angles. 27\% of the galaxies in our sample are misclassified, most of them being oblate galaxies classified as triaxial, or triaxial galaxies classified as prolate. One possible reason for this is that a small variation in the axis ratios $p$ and $q$ can lead to very different values of $T_{\rm Re}$ for nearly spherical systems (due to how $T_{\rm Re}$ is defined). In particular, if the reconstructed values of $p$ are different from the true ones, even though still consistent within the errors, they can lead to a large change in $T_{\rm Re}$. Moreover, the range in the shape parameters that we cover is determined by the minimum value of the axis ratio from the projected surface brightness/mass in the MGEs. For face-on views, it is very possible for the modelling to overestimate $q$, as seen in Fig. \ref{fig:pq_re}. In addition to that, as mentioned in the previous paragraphs, we take the DM component to be spherical, while we assume a triaxial stellar component. This means that the degeneracy between stellar mass and DM mass could influence the intrinsic shapes. 
Even though a high percentage (27\%) of galaxies are misclassified, we note that $T_{\rm Re}$ never changes as much as to have prolate galaxies misclassified as oblate galaxies or oblate galaxies as prolate.

For the fraction of orbits within 1 \re, we find that the cold, hot, and counter-rotating mass fractions are reconstructed qualitatively well, with median offset of 0.07, 0.06 and 0.02, respectively. We find that the model-derived fractions of the warm orbits are underestimated by 0.14. These results are in agreement with \cite{Jin2019}. The differences in the orbital distributions, however, can be at least partially attributed to a difference in measurement techniques. Even though we have averaged the instantaneous orbital distribution from the particle data using 3 different snapshots, these distributions are not the same as the time-averaged distributions we get from the models (which are calculated, by default, over 200 orbital periods, meaning that, especially at larger radii, the orbital distributions can be averaged over a Hubble time). Moreover, galaxy interactions or interactions with the environment can have an effect on the orbital distributions, and can lead to snapshot-averaged values being very different to what we would have if we calculated the circularity by freezing the potential, integrating the particle orbits in the potential, and then calculating the average values along the orbits (which is, however, computationally and time expensive). 

After checking the comparison between the model-derived properties and the true ones, we compare our results with the orbital properties of MAGPI galaxies derived using Schwarzschild dynamical modelling by D24. We find that our selected sample from the {\sc Eagle} galaxies reproduces the galaxy properties derived from observed MAGPI galaxies qualitatively well, showing similar trends in orbital fractions, triaxiality (although not individual axis ratios), and velocity anisotropy, although with a larger scatter in all of the relations explored. {\sc Eagle} galaxies, however, seem to have higher fractions of DM, compared to their MAGPI counterparts, and are less round in shape. The galaxies in the {\sc Eagle} and MAGPI samples are mostly non-oblate objects, which are likely to be radially anisotropic (\br $>$ 0), and to have mainly hot (box or radial) orbits, with few cold (short-axis tube) orbits.

To understand the connections between the different galaxy properties we analyse here and to understand the correlations we find for the galaxies in the {\sc Eagle} and MAGPI samples, we will now explore the connection between the merger history of our {\sc Eagle} galaxies and their intrinsic properties at $z \sim 0.3$. 

\subsection{Intrinsic properties and merger history}\label{sec:mergers}

We define mergers and merger histories in the same way as in \cite{Santucci2024}, but based on the merger history of galaxies at $z=0.27$: galaxies with ``no mergers'' are those that did not have a minor or major merger since $z=2$, with minor and major mergers being those with a stellar mass ratio between the second most massive and most massive progenitors between $[0.1-0.3]$ and $[0.3,1]$, respectively. 

\begin{figure} 
\centering
\includegraphics[width=\linewidth]{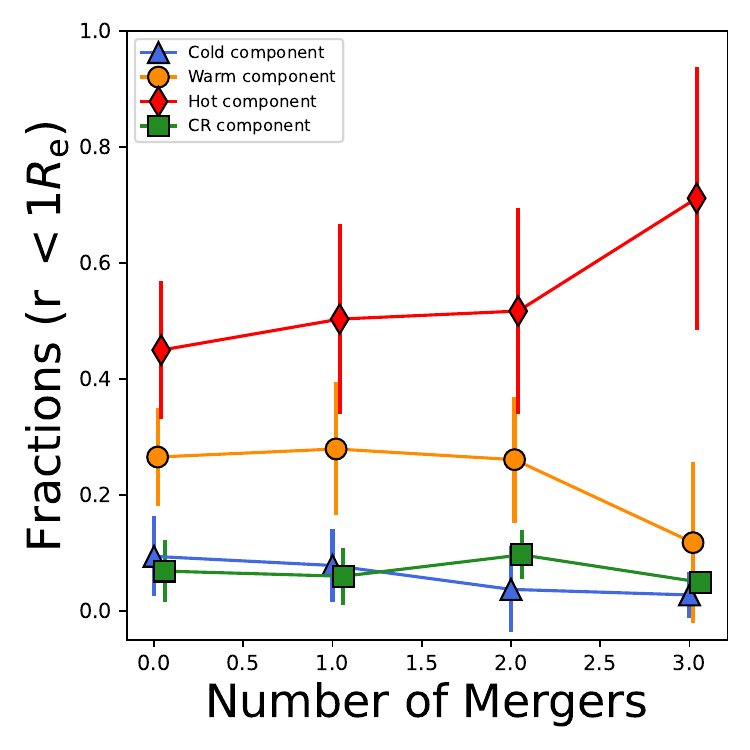}
\caption{ Medians (symbols connected by lines) of the fractions of each orbital component, as labelled, within 1 \re, as a function of the total number of minor and/or major mergers experienced by the galaxies in our sample from $z=2$ up to $z=0.27$. Errorbars show the 1-$\sigma$ scatter around the median. Note that we perturb the x-axis position of the symbols slightly to avoid overlapping errorbars. The fractions of hot orbits increase with increasing number of mergers, while the fractions of warm and cold orbits decrease with increasing number of mergers.}
\label{fig:fracs_nmergers}
\end{figure}

The majority of the galaxies in our sample ($\sim 69\%$) have experienced at least one merger (either minor or major) in their life, and 31\% of the galaxies have experienced 2 or more mergers, with the majority of these being minor mergers. 
In order to constrain the correlation between galaxy mergers and a galaxy's orbital properties, we show in Fig.~\ref{fig:fracs_nmergers} the median values of the orbital fractions within 1 \re~as a function of the total number of minor and/or major mergers each galaxy has undergone from $z=2$ up to $z =0.27$. We do not find any correlation between the fractions of counter-rotating orbits and the number of mergers. Even though the overall trends are consistent with constant, we see suggestions of possible correlations between the other fraction of orbits and the number of mergers: the fractions of hot orbits increase with increasing number of mergers, while the fractions of cold and warm orbits decrease, with galaxies that have experienced 3 or more mergers having $\sim 51\%$ more mass in hot orbits, $\sim 57\%$ less mass in warm orbits, and 66\% less mass in cold orbits than galaxies with $\le 1$ merger.  

\begin{figure} 
\centering
\includegraphics[width=\linewidth]{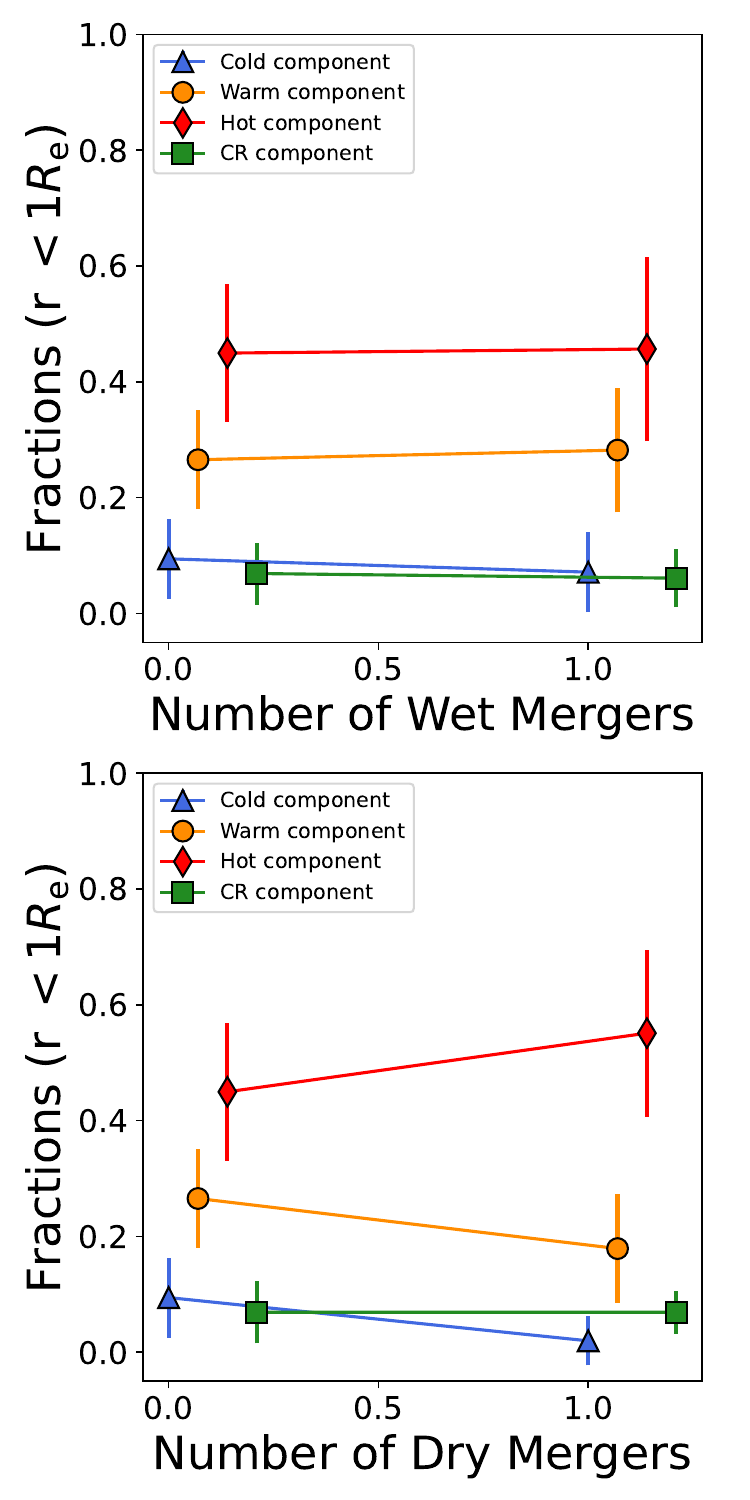}
\caption{ Medians (symbols connected by lines) of the fractions of each orbital component, as labelled, within 1 \re, as a function of the number of minor and/or major wet (top panel) and dry (bottom panel) mergers experienced by the galaxies in our sample from $z=2$ up to $z=0.27$. 
Errorbars show the 1-$\sigma$ scatter around the median. Note that we perturb the x-axis position of the symbols slightly to avoid overlapping errorbars. The fractions of hot orbits increase with increasing number of dry mergers, while the fractions of warm orbits decrease with increasing number of dry mergers. No change is found in the fraction of orbits as a function of wet mergers.}
\label{fig:fracs_wetdrys}
\end{figure}

Several theories have been proposed where different types of mergers can have different effects on a galaxy's structure and composition, leading to the formation of slow- or fast-rotating galaxies. For example, \citet{Cappellari2016} described how early-type galaxies can form through two main formation channels: fast-rotating galaxies start as star-forming discs and evolve through a set of processes dominated by gas accretion, bulge growth, and quenching. By comparison, slowly rotating galaxies assemble near the centre of massive haloes via intense star formation at high redshift \citep{Naab2009,Clauwens2018}, and evolve from a set of processes dominated by gas-poor mergers, resulting in more triaxial shapes. Gas-poor and gas-rich mergers can also have very different effects on the spin of galaxies (as shown by \citealt{Naab2014}, for example). More recently, \cite{Schulze2018} studied galaxies in the Magneticum simulation and found that at least 30\% of the population of slow rotators at $z \sim 0$ underwent the transition from fast to slow rotator after a significant merging event (with a mass ratio between 0.2 and 1). \citet{Lagos2022} used mock observations of galaxies with $M_{\star} > 10^{10} M_{\odot}$ from the {\sc Eagle} simulation and found that slow-rotating galaxies with triaxial or prolate shapes are usually formed as a consequence of major and minor mergers, with prolate shapes being driven by gas-poor mergers. Slow-rotating galaxies that only had very minor mergers are associated with triaxial shapes while those that formed in the absence of mergers are oblate.

To explore how orbital fractions are influenced by different types of mergers, we classify the mergers into gas-rich (wet) and gas-poor (dry) using the ratio of gas to stellar mass of the merger ($f_{\rm gas,merger}$, derived by \citealt{Lagos2018a}). The latter is the sum of the gas masses of the two merging galaxies divided by the sum of the stellar masses of the two galaxies. In this paper, we classify mergers with a gas fraction below $0.2$ as dry or gas-poor, and mergers with gas fractions above this value as wet or gas-rich \citep[as in][]{Santucci2024}. We find that the majority of the mergers experienced by the galaxies in our sample were wet mergers, with $\sim 35\%$ of the galaxies experiencing a dry merger - of these, 1/3 had undergone one major dry merger. We show the median values of the orbital fractions within 1 \re~as a function of the number of wet (top panel) and dry (bottom panel) minor and/or major mergers each galaxy has undergone from $z=2$ up to $z =0.27$ in Fig.~\ref{fig:fracs_wetdrys}. Note that each panel does not include galaxies that have experienced both wet and dry mergers. We find that the change in the fractions of hot and warm orbits that we see with number of mergers is driven by dry mergers, while no change in the fractions of orbits is found with increasing number of wet mergers.

This scenario and the changes in orbital fractions we find with number of mergers are consistent with the correlation we found between fraction of orbits and number of wet/dry mergers in \cite{Santucci2024}, where the effects of wet mergers on the orbital fractions are weak, but appear to accumulate with the number of wet mergers (so that no clear difference is found after one wet merger), while the effects of dry mergers are stronger, so that even one dry merger can increase the fraction of hot orbits and decrease the fractions of warm and cold orbits.

The high fraction of wet mergers can also explain the high number of triaxial galaxies and low number of prolate galaxies that we see in our sample. Previous studies found prolate systems with minor-axis rotation produced through collisionless equal-mass merger of disc galaxies \citep{Moody2014, RodriguezGomez2015}. \cite{Li2018prolate} found that almost all prolate galaxies in the Illustris simulation have experienced at least one dry merger (mainly major), with a radial merger orbit, and they find nearly no prolate galaxy with stellar mass $M_{\star} < 3 \times 10^{11} M_{\odot}$. On the other hand, \cite{Valenzuela2024} found that this relation is not as clear for galaxies in the Magneticum simulation. \cite{Lagos2022} found, for galaxies in the {\sc Eagle} simulation, that prolate shapes are driven by gas-poor mergers. \cite{Li2018prolate} also found that the majority of the galaxies they studied in the Illustris simulations had triaxial shapes between $z=2.32$ and $z=0.46$, due to the high number of wet mergers, which can often make the galaxy shapes irregular. \cite{Pulsoni2020} studied the shape of the stellar haloes of galaxies from the IllustrisTNG simulations and found a non-negligible fraction of galaxies with prolate shapes, in particular at $\log M_{\star}/M_{\odot} >11$, with the triaxiality increasing with increasing radius. They did not find any clear difference between the shapes of the stellar haloes (when looking at the outskirts) of fast- and slow-rotating galaxies.

For the future, it will be interesting to compare the galaxy populations at $z \sim 0.3$ and $z \sim 0$, to understand how galaxy shapes evolve with time, by comparing our results with observations, such as SAMI \citep{Santucci2022} and ATLAS$^{\rm 3D}$ \citep{Thater2023}, with shapes derived using triaxial Schwarzschild models.

\cite{Santucci2024} found good agreement between the orbital distributions of SAMI and {\sc Eagle} galaxies at $z \sim 0$. The fact that we also find good agreement between the orbital distribution of MAGPI and {\sc Eagle} galaxies is interesting on its own because it shows that, at least up to $z \sim 0.3$, the {\sc Eagle} simulations seem to reproduce the dynamical properties of the galaxies in the observed Universe qualitatively well.

\section{Summary}
We build Schwarzschild dynamical models for \Nsample~ MAGPI-like mock observations of {\sc Eagle} galaxies. We use the state-of-the-art {\sc{DYNAMITE}} code \citep{DYN2020}, allowing for fully triaxial Schwarzschild models, and make no assumptions on the orbital structure. We derive their intrinsic shapes and orbital distributions and compare them with the true values from the particle data. We then explore the distribution of the model-derived galaxy properties (DM fractions, intrinsic shapes, fraction of orbits, and velocity anisotropy) with stellar mass and as a function of other galaxy properties, and we compare our results with those derived for 22 MAGPI galaxies by D24.

For our {\sc Eagle} sample we find that:
\begin{itemize}

\item the enclosed DM mass within 1 \re~ is well reproduced when the DM has a distribution that is well represented by an NFW profile. However, enclosed DM can be underestimated in the other cases, with a median offset of 48\%;

\item  73\% of the galaxies in our sample have model-derived shapes consistent with their intrinsic ones, with a median offset in triaxiality of 0.07. Even though the exact values of individual galaxies can be off, in studying a bulk population we are able to get a statistical view of their shapes;

\item the cold, hot and counter-rotating mass fractions are reconstructed qualitatively well, with median offsets of 0.07, 0.06, and 0.02, respectively, while the model-derived fractions of warm orbits have a median offset of 0.14;

\item our selected sample from the {\sc Eagle} galaxies tends to reproduce the trends found from observed MAGPI galaxies (by D24), showing similar trends in orbital fractions, triaxiality and velocity anisotropy, although with a larger scatter in all of the relations explored. {\sc Eagle} galaxies, however, seem to have higher fractions of DM, compared to their MAGPI counterparts, in particular at $\log M_{\star}/M_{\odot} < 11.2$. The galaxies in the {\sc Eagle} and MAGPI samples are mostly non-oblate ($T_{\rm Re} \geq 0.3$) objects, which are likely to be radially anisotropic (\br $>$ 0), and to have mainly hot (box or radial) orbits, with few cold (short-axis tube) orbits;

\item over half of the galaxies in our sample ($\sim 69\%$) have experienced at least one merger (either minor or major) in their life. This seems to lead to an increase in the fractions of hot orbits with increasing number of mergers, and a slight decrease in the fractions of cold and warm orbits with increasing number of mergers. The effect, however, is cumulative and seems to have some impact only on the orbital fractions of galaxies with 2 or more mergers;

\item the majority of the mergers experienced by the galaxies in our sample were wet mergers - with 35\% of the galaxies in our sample experiencing a dry merger - explaining the weak, but cumulative effects of mergers on the fraction of orbits.
\end{itemize}

Given the array of simulations present in the MAGPI theoretical library, it would be interesting to see how our results compare to properties derived from mock observations of different simulations. This comparison could also help to understand whether the differences we see between simulations and observations come from the different "recipes" used to create them, or from the underlying physics.

\section*{Acknowledgements}
We warmly thank the reviewer for their comments which helped improve this work.
We wish to thank the ESO staff, and in particular the staff at Paranal Observatory, for carrying out the MAGPI observations. MAGPI targets were selected from GAMA. GAMA is a joint European-Australasian project based around a spectroscopic campaign using the Anglo-Australian Telescope. GAMA was funded by the STFC (UK), the ARC (Australia), the AAO, and the participating institutions. GAMA photometry is based on observations made with ESO Telescopes at the La Silla Paranal Observatory under programme ID 179.A-2004, ID 177.A-3016. The MAGPI team acknowledge support by the Australian Research Council Centre of Excellence for All Sky Astrophysics in 3 Dimensions (ASTRO 3D), through project number CE170100013. This work was performed on the OzSTAR national facility at Swinburne University of Technology. The OzSTAR program receives funding in part from the Astronomy National Collaborative Research Infrastructure Strategy (NCRIS) allocation provided by the Australian Government, and from the Victorian Higher Education State Investment Fund (VHESIF) provided by the Victorian Government. 
GS thanks Luca Cortese and Bodo Ziegler for their suggestions, which improved the manuscript.
GS and KH acknowledge funding from the Australian Research Council (ARC) Discovery Project DP210101945. GS acknowledges funding from the UWA Research Collaboration Awards 2022. CL has received funding from the ARC Centre of Excellence for All Sky Astrophysics in 3 Dimensions (ASTRO 3D) through project number CE170100013. AP acknowledges support from the Hintze Family Charitable Foundation. AFM acknowledges support from RYC2021-031099-I and PID2021-123313NA-I00 of MICIN/AEI/10.13039/501100011033/FEDER, UE, NextGenerationEU/PRT. CL, JTM and CF are the recipients of ARC Discovery Project DP210101945. LMV acknowledges support by the German Academic Scholarship Foundation (Studienstiftung des deutschen Volkes) and the Marianne-Plehn-Program of the Elite Network of Bavaria. SMS acknowledges funding from the Australian Research Council (DE220100003).
Parts of this research were conducted by the Australian Research Council Centre of Excellence for All Sky Astrophysics in 3 Dimensions (ASTRO 3D), through project number CE170100013.
\section*{Data Availability}

The data underlying this article are available through \href{http://icc.dur.ac.uk/Eagle}{http://icc.dur.ac.uk/Eagle}. Data products created for this article are available upon request to the \sendemail{giulia.santucci@uwa.edu.au}{Eagle orbital distributions: data request}{corresponding author}.

Measurements from Schwarzschild models of MAGPI galaxies are available by contacting \sendemail{richard.mcdermid@mq.edu.au}{MAGPI Schwarzschild models: data request}{
Richard McDermid.}



\bibliographystyle{mnras}
\bibliography{references} 




\appendix
\section{Instantaneous and time-averaged orbital distributions}\label{app:frozen_pot}
\begin{figure*} 
\centering
\includegraphics[width=\linewidth]{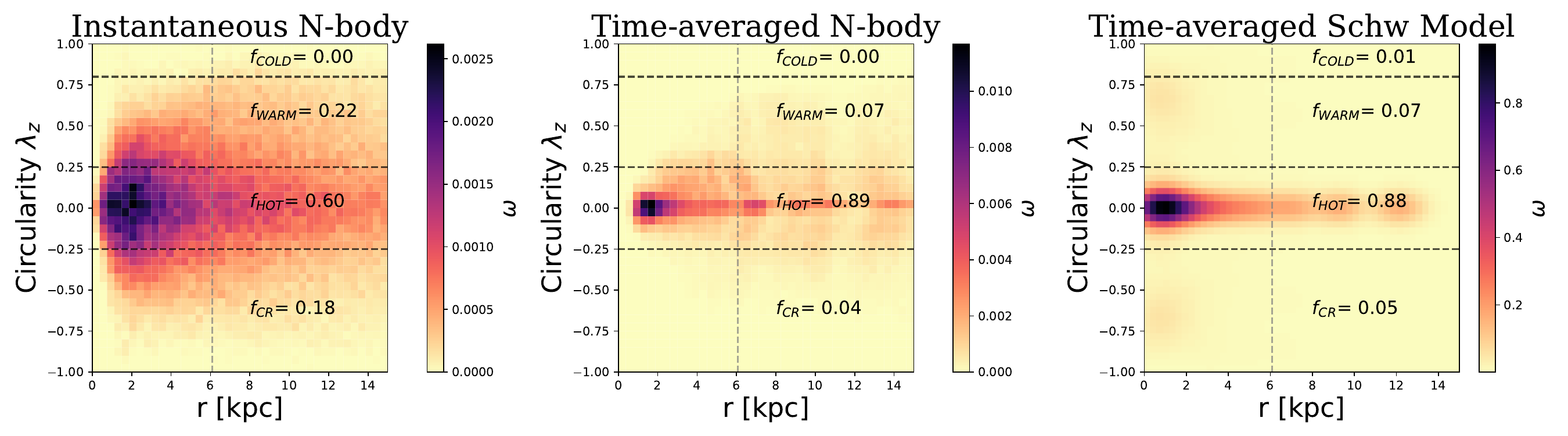}
\caption{Example galaxy {\sc Eagle} 9498015: orbit circularity $\lambda_z$ versus radius (in kpc). The colour indicates the density, $w$, of the orbits on the phase space, and the three horizontal black dashed lines indicate $\lambda_z$ = 0.8, $\lambda_z$ = 0.25 and $\lambda_z$ = -0.25, dividing the orbits into four regions (cold, warm, hot and counter-rotating orbits). Fractions of mass within 1\re~in each component are annotated in the figures. The vertical grey dashed line is located at 1 $R_{\rm e}$. Left panel shows the instantaneous orbital distribution of the stellar particles in the N-body model after 10 Gyr; middle panel show the time-averaged orbital distribution of the stellar particles in the N-body model; right panel shows the time-averaged orbital distribution derived from the dynamical modelling. Time-averaging helps to reconstruct the orbital distribution in a similar manner to what is derived by {\sc{DYNAMITE}}. }
\label{fig:frozen_orbital_dist}
\end{figure*}

In order to justify the time-averaging of the orbital distributions within the simulation for comparison with our model-derived results from  {\sc DYNAMITE}, we create an N-body model of the {\sc{Eagle}} galaxy 9498015 using the Schwarzschild-derived best-fit potential. Following the approach of \cite{Zhu2022halo}, we keep the potential fixed (similar to what is being done using {\sc{DYNAMITE}}) and allow the stellar particles to evolve within it. We then measure the instantaneous orbital distributions of the stellar particles at different times, as well as the time-averaged distribution. The time-average distribution is calculated from time-averaged positions and velocities of the stellar orbits. 

In Fig. \ref{fig:frozen_orbital_dist} we show the comparison between the instantaneous orbital distribution from the N-body model after 10 Gyr (left panel), the time-averaged distribution of the particles in the N-body model (middle panel), and the orbital distribution from the best-fit Schwarzschild dynamical model (right panel). We also indicate the values of the orbital fractions within 1 $R_{\rm e}$ for each component. 

The instantaneous orbital distribution for the N-body model is, as expected, very close to the true orbital distribution (shown in Fig. \ref{fig:circularity}), and the time-averaged distribution aligns closely with what we obtain from the Schwarzschild models, with very similar orbital fractions within 1 $R_{\rm e}$. This suggests that time averaging is necessary to compare orbital fractions derived from simulations with the Schwarzschild model-derived values in a consistent manner.

\section{Fraction of orbits from instantaneous distributions}\label{app:orbits}

\begin{figure*} 
\centering
\includegraphics[width=\linewidth]{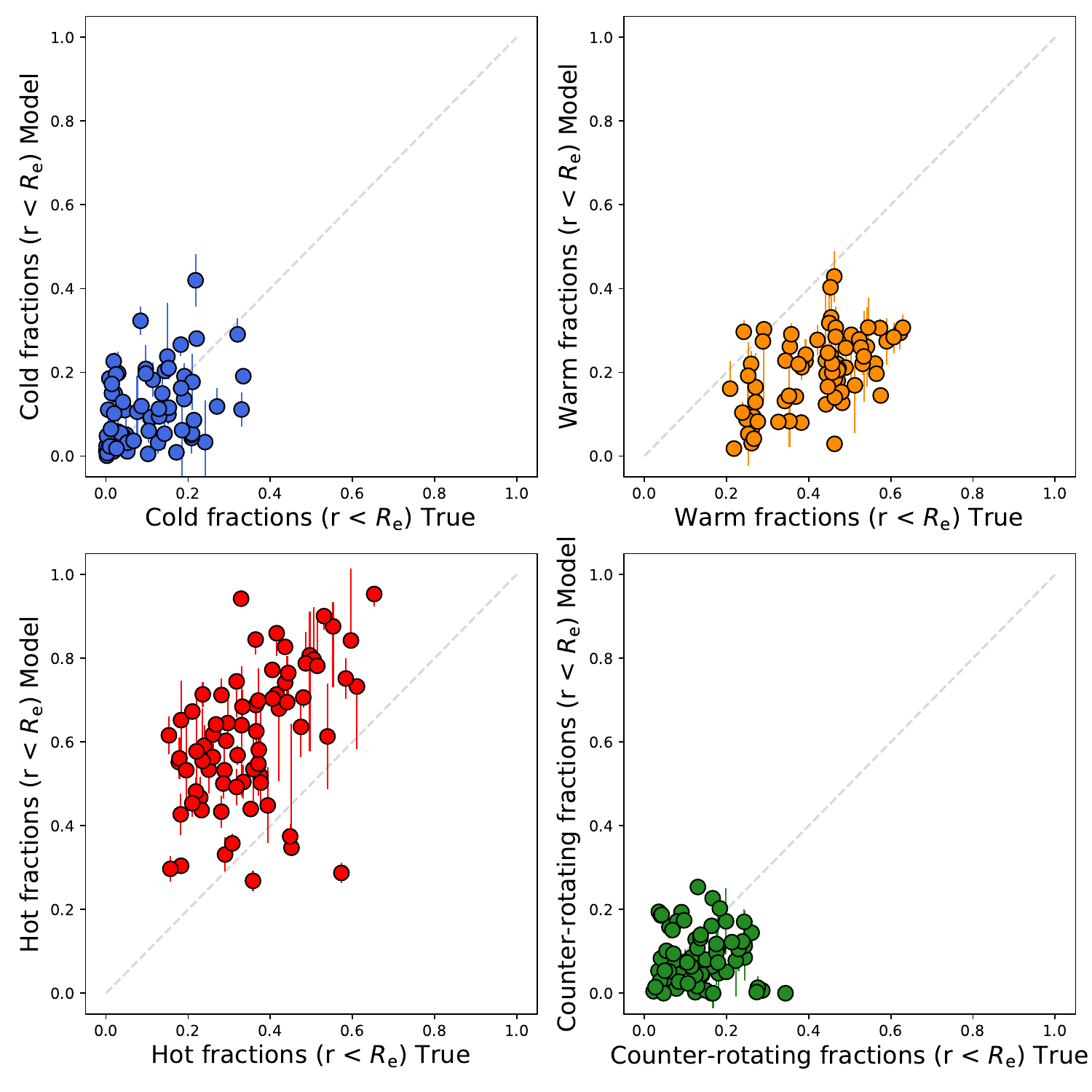}
\caption{One-to-one comparison of the true and model-recovered fraction of the cold (in blue), warm (in orange), hot (in red), and counter-rotating (in green) orbits. The true values are calculated from the instantaneous orbital distribution. We find that, for $\lambda_z$, the comparison between the model-derived and true fractions of cold and counter-rotating orbits are in general agreement, within the errors. However, model-derived fractions of hot orbits are generally higher than the instantaneous fractions, while the model-derived fractions of warm orbits tend to be lower than the instantaneous ones.}
\label{fig:orbits_inst}
\end{figure*}

We show in Fig. \ref{fig:orbits_inst} the one-to-one comparison of the instantaneous and model-recovered fractions of the cold, warm, and hot orbits. We find that, for $\lambda_z$, the comparison between the model-derived and true fractions of cold and counter-rotating orbits are generally in agreement, within the errors. However, model-derived fractions of hot orbits are higher than the instantaneous fractions (with a median offset of 0.28), while the model-derived fractions of warm orbits tend to be lower than the instantaneous ones (with a median offset of 0.23. As mentioned in Sec. \ref{section:orbits}, this is not surprising since the model-derived orbital distributions are time-averaged.

\bsp	
\label{lastpage}
\end{document}